\documentclass[conference]{IEEEtran}

\IEEEoverridecommandlockouts
\usepackage{cite}
\usepackage{amsmath,amssymb,amsfonts}
\usepackage{algorithmic}
\usepackage{graphicx}
\usepackage{mathptmx}
\usepackage{textcomp}
\usepackage{xcolor}
\usepackage{subfigure}
\usepackage{booktabs}
\usepackage{blindtext}
\usepackage{diagbox}

\def\BibTeX{{\rm B\kern-.05em{\sc i\kern-.025em b}\kern-.08em
    T\kern-.1667em\lower.7ex\hbox{E}\kern-.125emX}}

\begin{document}

\title{Taylor Modeling and Comparative Research Containing Aspect-Ratio Dependent Optimization of Three-Dimensional H$k$ Superjunction MOSFETs 



\author{
    Zhentao Xiao$^{1,2*}$, Haimeng Huang$^{1}$, Zonghao Zhang$^{1}$, Chenxing Wang$^{1,2}$
    \\
   \small \textit{$^1$University of Electronic Science and Technology of China, Chengdu 610054, China} \\
    \small \textit{$^2$University of Glasgow, Glasgow, G12 8QQ, United Kingdom} \\
    *Corresponding author: 2839952X@student.gla.ac.uk
}
}

\maketitle

\begin{abstract}

This paper presents a comprehensive study on aspect-ratio dependent optimization for specific on-resistance of three-dimensional high-\emph{k} superjunction MOSFETs. The research introduces a Taylor modeling method, overcoming the computational limitations of the Bessel method. It also employs the Chynoweth model for more accurate breakdown voltage determination. The study provides a comparative analysis of four different superjunction structures, across five aspects: electric field, impact ionization integral, aspect ratio dependent optimization, charge imbalance effect and temperature. The findings offer valuable insights for the manufacturing guidance of superjunction structure selection.
\par\

\end{abstract}


\section{Introduction}
\begin{figure}[t]
 \subfigure[]{
    \includegraphics[scale=0.3]{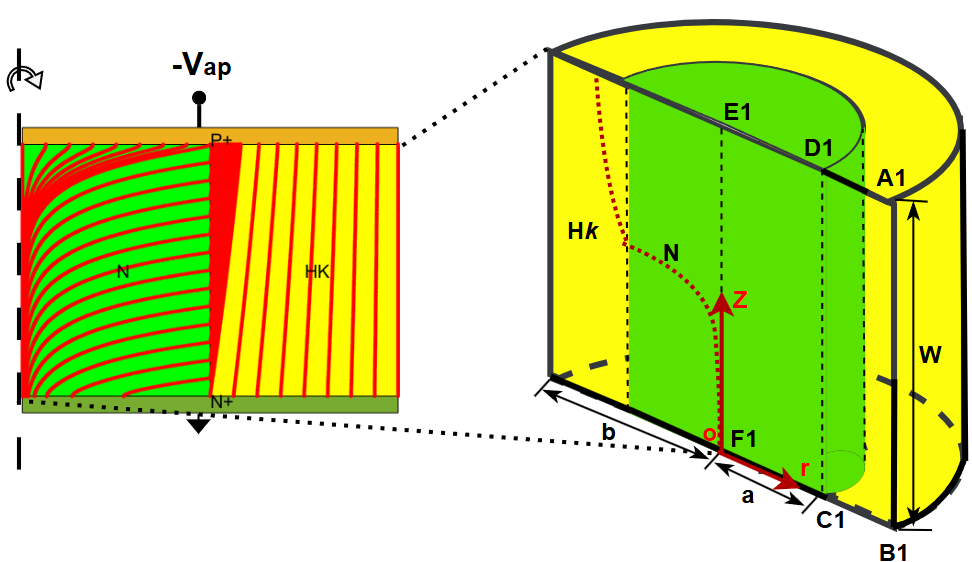}
    }
    
     \subfigure[]{
    \includegraphics[scale=0.36]{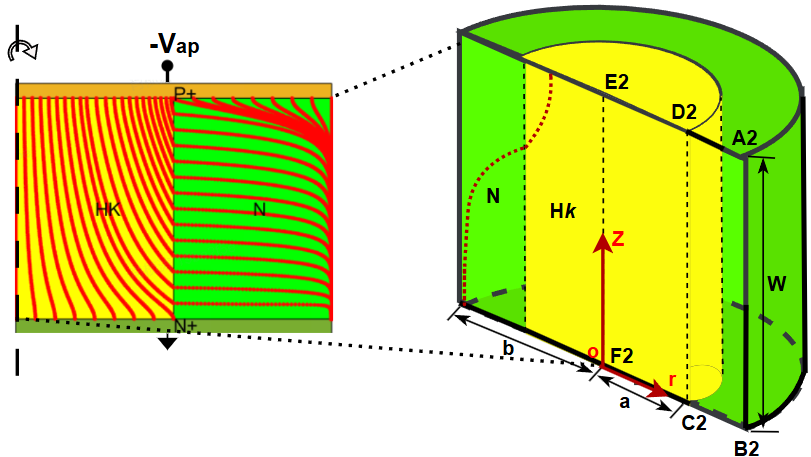}
    }
    \vspace{-3mm}
    \caption{Structure and E-field lines of (a) 3DH$k$case1 and (b) 3DH$k$case2. The E-field lines are calculated and drawn by MATLAB using the methodology in \cite{EFL}.}
    \label{structure1}
\end{figure}

In the field of power devices, the breakdown voltage (BV) and the specific on-resistance (\emph{R}$_{\rm \textbf{on,sp}}$) are the two critical parameters for assessing the quality of a device. However, it has been discovered that BV and \emph{R}$_{\rm \textbf{on,sp}}$ are interdependent, a constraint known as the 'Silicon Limit'. The introduction of the Superjunction (SJ) structure has greatly alleviated this constraint\cite{cxb}\cite{siliconlimit}\cite{napoli}\cite{cxb1998}\cite{chen2002}, allowing devices to achieve higher BV and lower \emph{R}$_{\rm \textbf{on,sp}}$ through the lateral steering of electric field lines (E-field lines). Due to the charge imbalance and pronounced JFET effect in conventional SJs (C-SJ)\cite{hhm202203}\cite{3DJFET}\cite{zwt}, the High-$k$ material (H$k$) has been proposed to replace the P-pillar to form the high-$k$ superjunction (H$k$-SJ)\cite{2013}\cite{methodology}\cite{EFL}\cite{HK}. Beyond this, driven by the need for process optimization and further performance enhancement, both H$k$-SJs and C-SJs are evolving towards a three dimensional (3D) direction \cite{hhm202203}\cite{3DJFET}\cite{zwt}\cite{2013}\cite{3D}. However, compared to the research on 3D C-SJs, the research on 3D H$k$-SJs is scarce and not in-depth. Generally speaking, the 3D H$k$-SJ is divided into two categories\cite{2013}, as shown in Fig. \ref{structure1}(a) and (b), namely 3DH$k$case1 and 3DH$k$case2, which differ in the position of the H$k$ layer and the lateral steering of E-field are also shown.

Through literature review, the cutting-edge 3D H$k$-SJ research still has the following shortcomings: 1. The E-field modeling of 3D H$k$-SJs is limited to the Bessel method\cite{2013}, which is computationally intensive and time-consuming. 2. The optimization of 3D H$k$-SJs is not comprehensive, with no optimization dependent on the aspect ratio. 3. Only the Fulop impact ionization integral model is used\cite{2013}, with no application of the Chynoweth precise model, leading to inaccurate BV \cite{hhm202203}. 4. The research on 3D H$k$-SJs is relatively isolated, with no comparative studies between 2D H$k$-SJs (2DH$k$) and different types of 3D H$k$-SJs, resulting in poor manufacturing guidance.

Considering the aforementioned reasons, this article proposes a Taylor modeling method for 3DH$k$case2 in Section II and uses the Taylor method and Chynoweth model to complete the aspect ratio dependent optimization. Section III will consider a comprehensive comparison and analysis of the advantages and disadvantages of 3D C-SJ, 2DH$k$, 3DH$k$case1, and 3DH$k$case2, comparing from five aspects: E-field, impact ionization integral, aspect ratio dependent optimization, charge imbalance, and temperature. Finally, in Section IV, we will place the 3D H$k$-SJ structure within MOSFETs and complete a comparative analysis of electrical characteristics. Section V will summarize the entire article.

\section{Analytic E-Field Modeling of 3DH$k$-SJ}
\label{section2}
\subsection{Error Correction of Bessel Method}

Before presenting the Taylor modeling approach, we first correct the error in the potential distribution expression ($V_{\mathrm{H}k}$) in the H$k$ region of 3DH$k$case2 from \cite{2013}. First, the Poisson equation in 3D cylindrical coordinates is given by
\begin{figure}[!h]
\centering
    \subfigure[]{\includegraphics[width=0.8\linewidth]{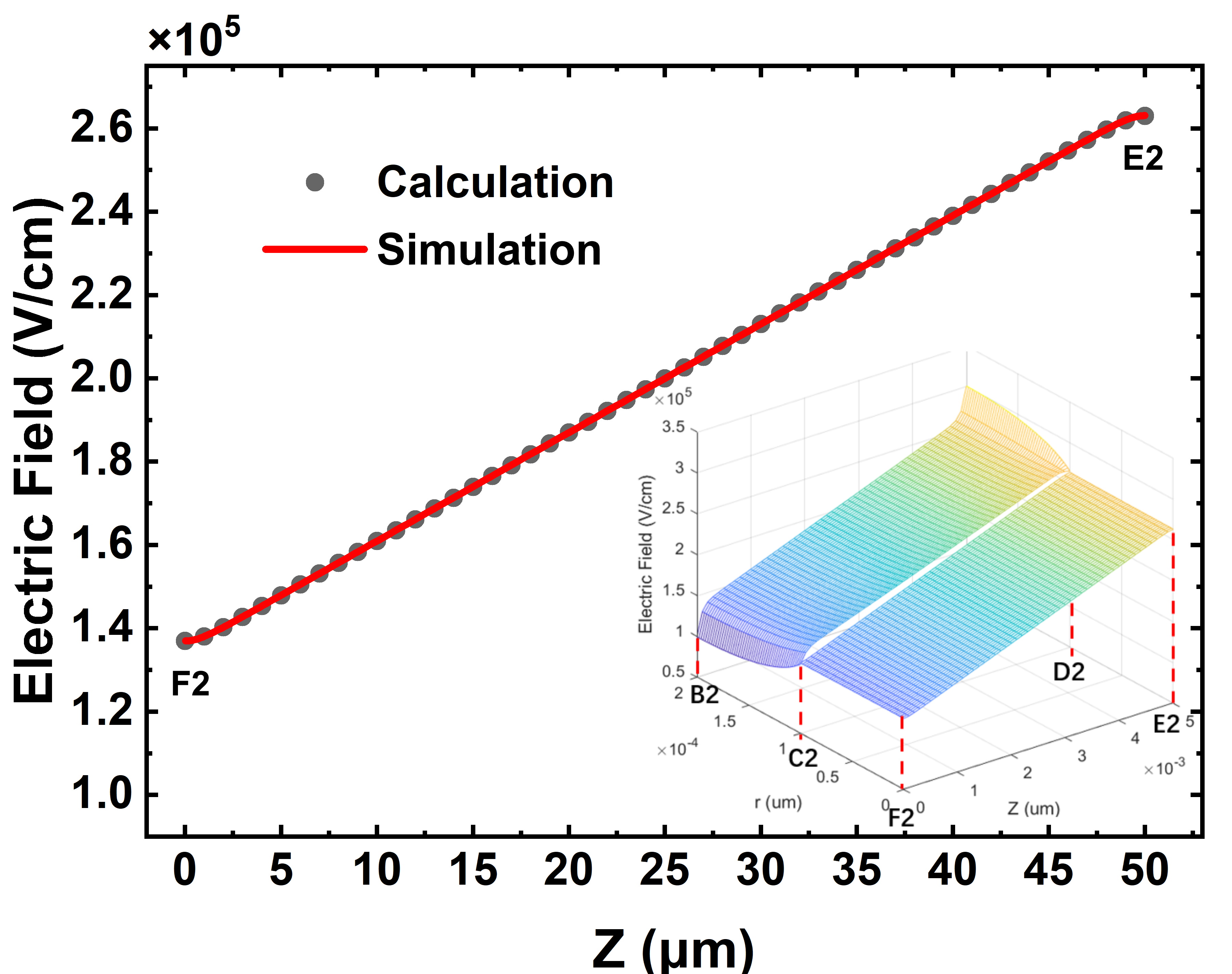}
    }
    \subfigure[]{\includegraphics[width=0.8\linewidth]{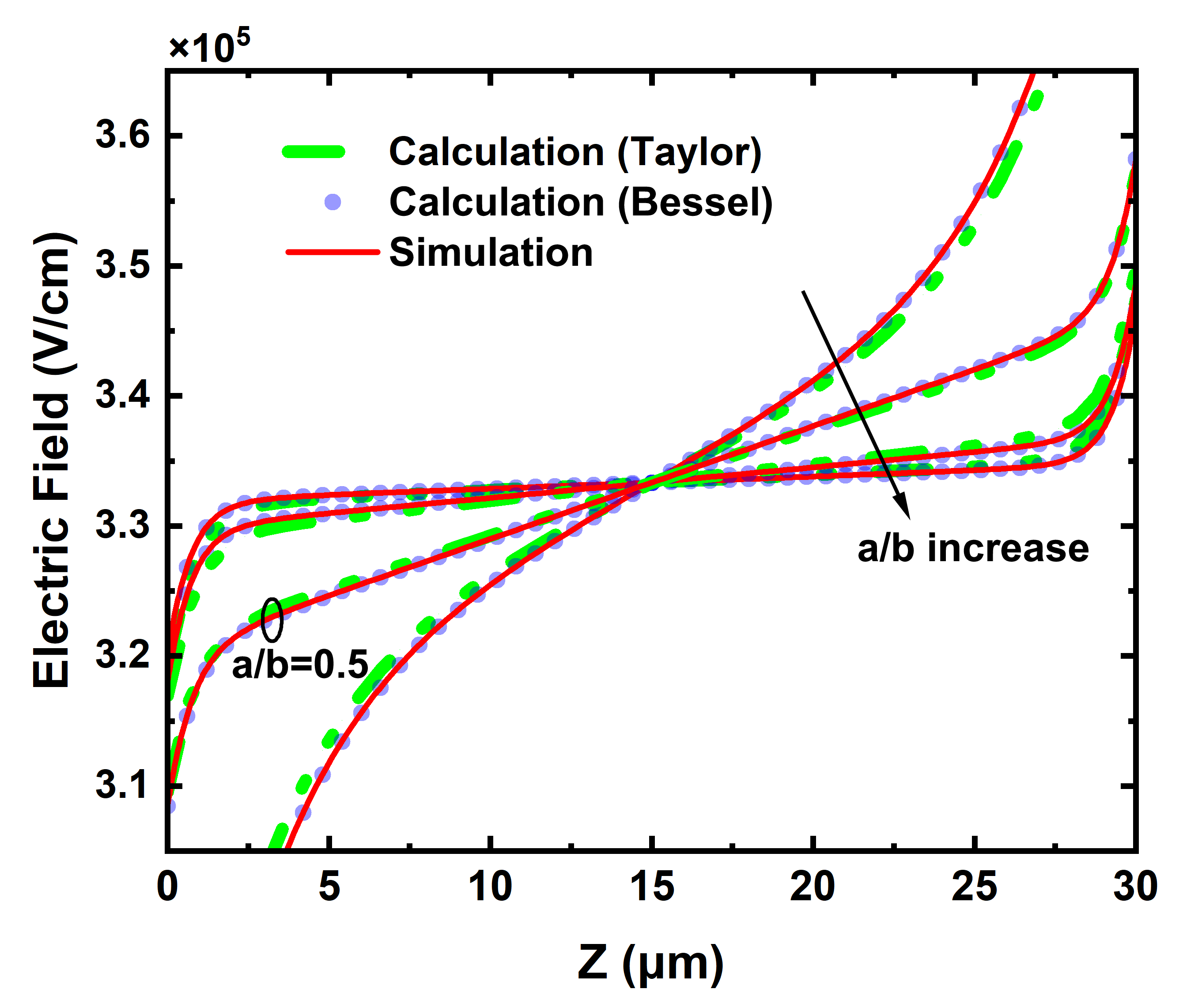}
    }
    \subfigure[]{\includegraphics[width=0.8\linewidth]{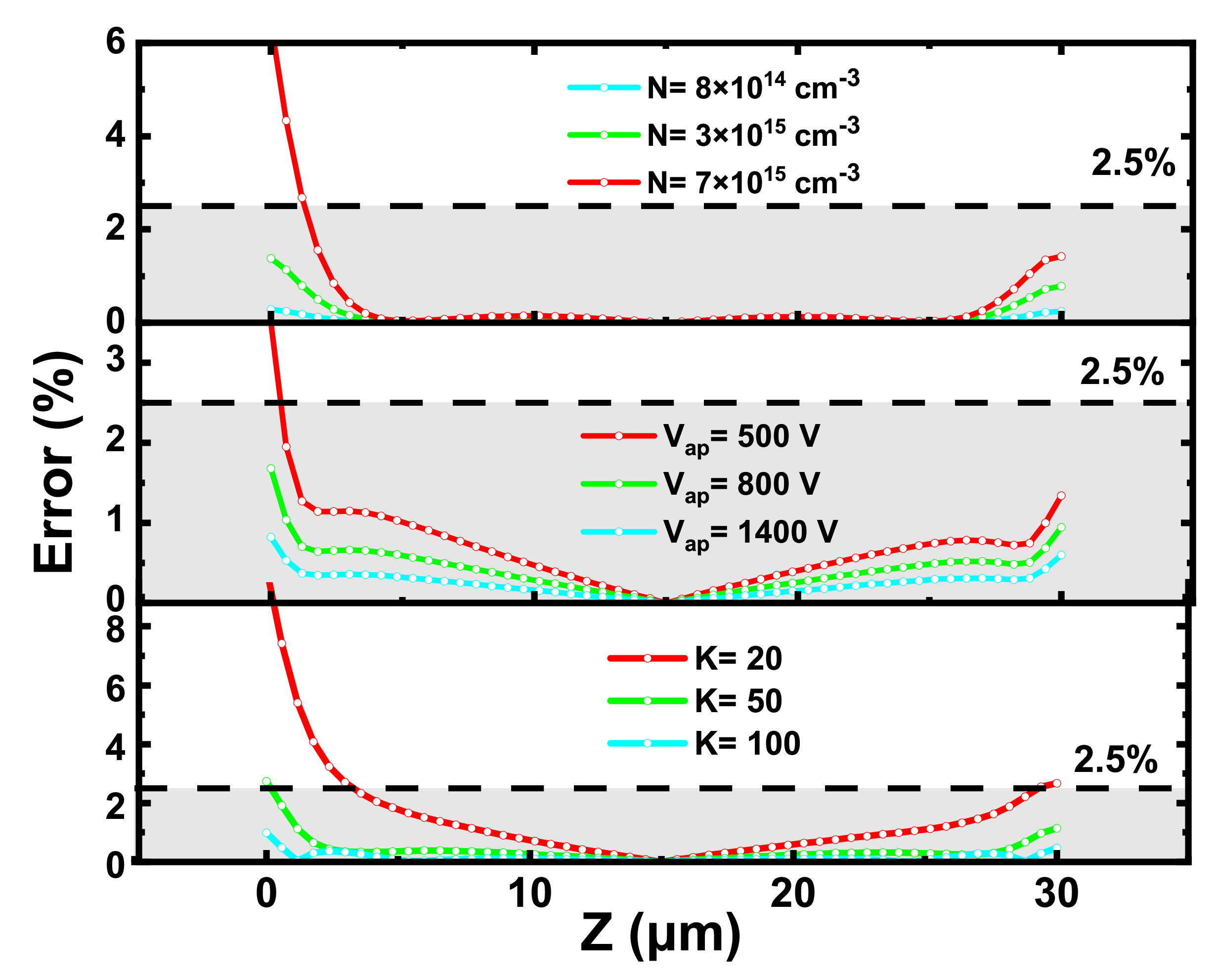}
    }
    \vspace{-2.1mm}
    \caption{(a) The comparison between the calculation using equation (4) and simulation result along $F_{2}E_{2}$, with 3D depiction of entire E-field distribution under same condition. (b) The comparison between Bessel\cite{2013}, Taylor method and simulation under different a/b at r=b. (c) The error of Taylor method along $A_{2}$$B_{2}$ at r = b under different conditions. Error= abs(100\%×[E(Taylor)-E(Bessel)]/E(Bessel)).}
    \label{Taylor}
\end{figure}
\begin{equation}
\begin{cases}
    \nabla^2 V_{\rm S}(r,z,\theta)=-\frac{q N}{\epsilon_{ \rm S}},\\
    \nabla^2 V_{\mathrm{H}k}(r,z,\theta)=0.\\
    \end{cases}
\end{equation}
The equations above can be expanded and simplified as followes due to the symmetry,
\begin{equation}
\begin{cases}
\frac{\partial^{2}V_{\rm S}}{\partial r^{2}}+(\frac{\partial V_{\rm S}}{\partial r})\frac{1}{r}+\frac{\partial^{2}V_{\rm S}}{\partial z^{2}}=-\frac{q N}{\epsilon_{\rm S}} ,  \\
\frac{\partial^{2}V_{\mathrm{H}k}}{\partial r^{2}}+(\frac{\partial V_{\mathrm{H}k}}{\partial r})\frac{1}{r}+\frac{\partial^{2}V_{\mathrm{H}k}}{\partial z^{2}}=0 .  
\end{cases}
\end{equation}

For 3DH$k$case2, the boundary conditions can be written as
\begin{equation}
\begin{cases}   
V_{\rm S}\left( z=0 \right)=V_{\mathrm{H}k}\left( z=0 \right)=0,\\
V_{\rm S}\left( z=W \right)=V_{\mathrm{H}k}\left( z=W \right)=-V_{\rm ap},\\
\epsilon_{\rm S}\left.\frac{\partial V_{\rm S}}{\partial r}\right|_{r=a}=\epsilon_{\mathrm{H}k}\left.\frac{\partial V_{\mathrm{H}k}}{\partial r}\right|_{r=a},\\
V_{\rm S}\left(r=a\right)=V_{\mathrm{H}k}\left(r=a\right),\\
\left.\frac{\partial V_{\rm S}}{\partial r}\right|_{r=b}=\left.\frac{\partial V_{\mathrm{H}k}}{\partial r}\right|_{r=0}=0,
\end{cases}
\end{equation}
where, $V_{\rm ap}$ means the applied voltage as shown in Fig.\ref{structure1}. Finally, we can solve the potential distribution of H$k$ region from equations (2) and (3):
\begin{equation}
V_{\mathrm{H}k}=-\frac{V_{\rm ap}}{W}z+\frac{4NW^{2}q}{\pi^{3}\varepsilon_{\mathrm{H}k}}\cdot\sum_{n= \rm odd}\frac{I_{0}(\frac{n\pi r}{W})\sin(\frac{n\pi z}{W})}{n^{3}[\frac{\varepsilon_{\rm S}}{\varepsilon_{\mathrm{H}k}}I_{0}(\frac{n\pi a}{W})-\frac{\beta_{\rm n}}{\alpha_{\rm n}}I_{1}(\frac{n\pi a}{W})]}
\end{equation}
where, 
$\alpha_{\rm n}=I_{1}(\frac{n\pi a}{W})K_{1}(\frac{n\pi b}{W})-I_{1}(\frac{n\pi b}{W})K_{1}(\frac{n\pi a}{W})$ and $\beta_{\rm n}=I_{0}(\frac{n\pi a}{W}) K_{1}(\frac{n\pi b}{W})+I_{1}(\frac{n\pi b}{W})K_{0}(\frac{n\pi a}{W})$, respectively. 
 Fig. \ref{Taylor}(a) shows a comparison between the results calculated from equation (4) and the MEDICI (TCAD software) simulation results. The excellent agreement between the two confirms the accuracy of equation (4).

\subsection{E-Field's Taylor Method Modeling}
For the Taylor modeling method, we also use equation (2) and the boundary conditions in equation (3). Additionally, we perform a Taylor expansion of $V_{\rm S}(r,z)$ and $V_{\mathrm{H}k}(r,z)$ at $r$ = b  and $r$ = 0, respectively, retaining only the first three terms of the expansion. They are found as:
\begin{equation}
\begin{cases}V_{\rm S}(r,z)=V_{\rm S}(b,z) +\left.\frac{\partial V_{\rm S}(b,z)}{\partial r}\right|_{r=b}\cdot(r-b)\\\quad\quad\quad\quad +\left.\frac{\partial^{2}V_{\rm S}(b,z)}{\partial r^{2}}\right|_{r=b}\cdot\frac{(r-b)^{2}}{2},\\V_{\mathrm{H}k}(r,z)=V_{\mathrm{H}k}(0,z)+\left.\frac{\partial V_{\mathrm{H}k}(0,z)}{\partial r}\right|_{r=0}\cdot r\\\quad\quad\quad\quad +\left.\frac{\partial^{2}V_{\mathrm{H}k}(0,z)}{\partial r^{2}}\right|_{r=0}\cdot\frac{r^{2}}{2}.\end{cases}
\end{equation}

Then, through the calculation, $V_{\rm S}(b,z)$ can be found as
\begin{align}
V_{\rm S}(b,z)=(\frac{2}{\lambda^{2} T_{\rm d}^{2}}-1+\frac{b^{2}}{{2}T_{\rm d}^{2}})[A\exp(\lambda z)+B\exp(-\lambda z)]\nonumber \\   -\frac{T_{\rm e}^{2}}{T_{\rm d}^{2}} \frac{q N}{\epsilon_{\rm S}} z^{2}+Dz+E,
\end{align}

where $T_{\rm c}$, $T_{\rm d}$, and $T_{\rm e}$ are dimensioned constants with units of $\mu$m, $T_{\rm c}^{2}=-\frac{1}{2}\left[\frac{\epsilon_{\rm S}}{\epsilon_{\mathrm{H}k}} a (a-b)-1.42(a-b)^{2}\right]$, $T_{\rm d}^{2}=\frac{1}{2}\left[a^{2}-\frac{\epsilon_{\rm S}}{\epsilon_{\mathrm{H}k}} a (a-b)\right]$, and 
$\frac{1}{T_{\rm e}^{2}}=\frac{2}{T_{\rm d}^{2}}+\frac{1}{T_{\rm c}^{2}}$. Above three are structural constants which only depend on the structure parameters such as a and b. $\lambda$ is the root of characteristic equation of differential equation during the calculation and $\lambda = \left ( T_{\mathrm{e}} \cdot \sqrt{1-\left ( b/T_{\mathrm{d} } \right )^2 /2 }  \right ) ^{-1}$. The coefficients $A$, $B$, $D$ and $E$ are found as
\begin{equation}
\left\{\begin{aligned}A &= \frac{T_{\rm e}^2 q N / \epsilon_{\rm S}}{\exp(W\lambda) + 1}, \\ B &= \frac{T_{\rm e}^2 q N / \epsilon_{\rm S} \cdot \exp(W\lambda)}{\exp(W\lambda) + 1}, \\ D &= -\frac{T_{\rm d}^2 V_{\rm ap} - T_{\rm e}^2 q N W^2 / \epsilon_{\rm S}}{T_{\rm d}^2 W}, \\ E &= -\frac{T_{\rm e}^2 q N / \epsilon_{\rm S} \cdot \left(-2T_{\rm d}^2 \lambda^2 + b^2 \lambda^2 + 4\right)}{2T_{\rm d}^2 \lambda^2}.\end{aligned}\right.
\end{equation}

\normalsize{
\par The distribution of the electric field at  r = b  is derived by differentiating equation (6) and is given as
}
\normalsize{
\begin{align}
E_{\rm S}(b,z)
= -\left(\frac{2}{\lambda^2 T_{\rm d}^2} - 1 + \frac{b^2}{2 T_{\rm d}^2}\right) \left[A \lambda \exp{(\lambda z)} - B \lambda \exp{(-\lambda z)}\right]
\notag
\\+ 2 \frac{T_{\rm e}^2}{T_{\rm d}^2} \frac{q N}{\epsilon_{\rm S}} z - D.
\end{align}
}

\begin{figure}[!t]
    \centering
    \includegraphics[width=0.4\linewidth]{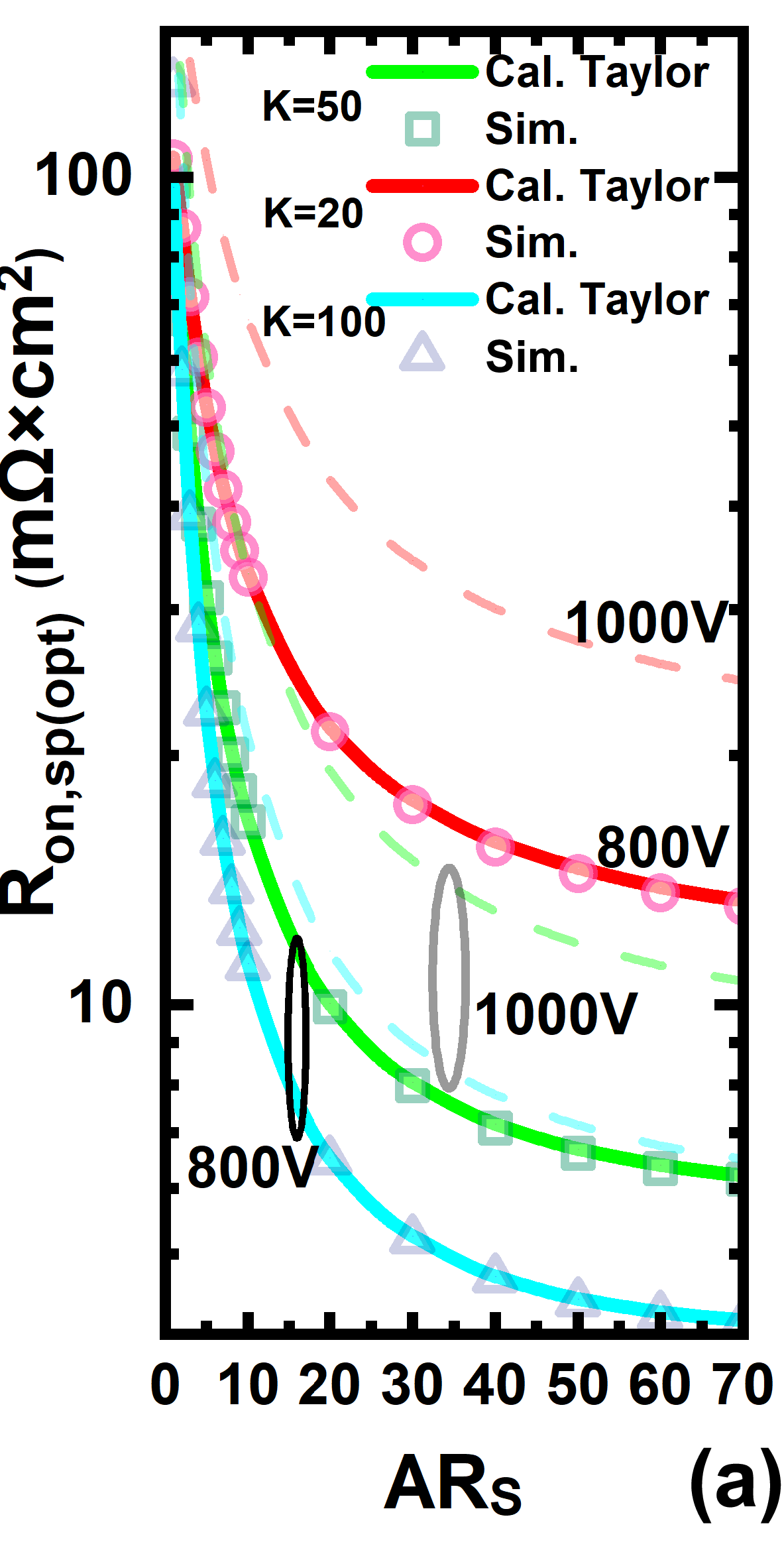}
    \hspace{-1mm}
    \includegraphics[width=0.4\linewidth]{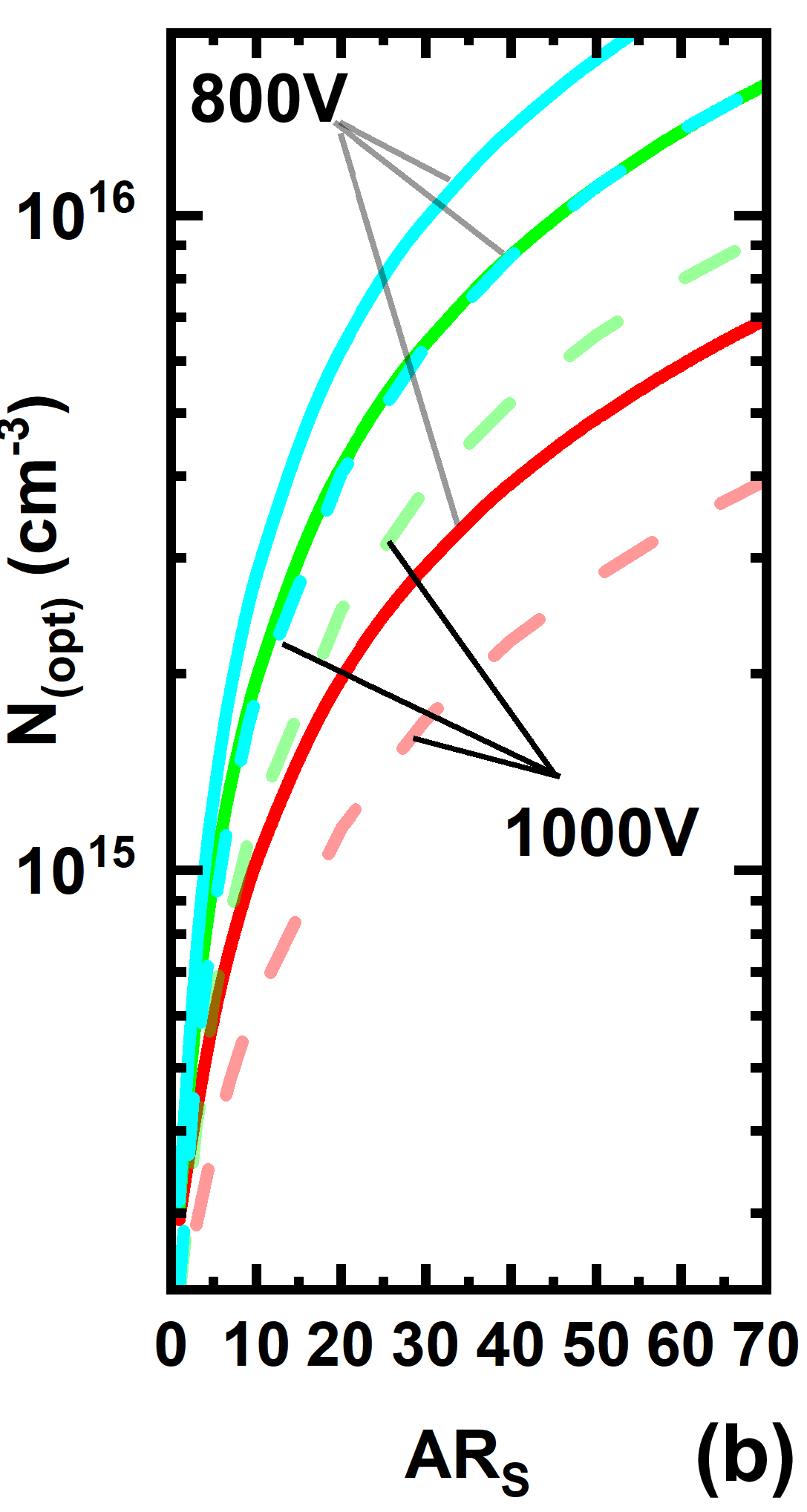}  
    \includegraphics[width=0.4\linewidth]{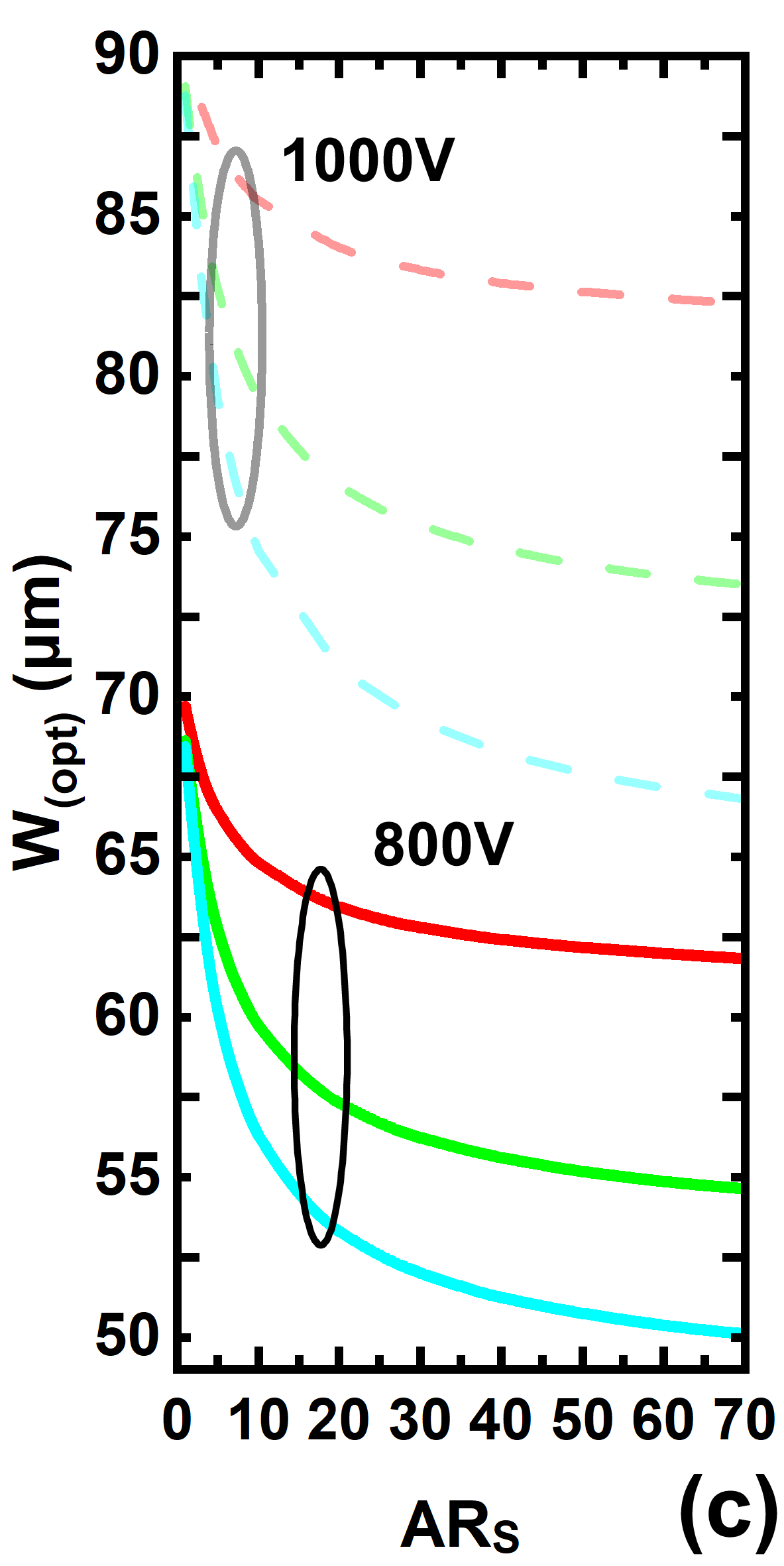}  
    \includegraphics[width=0.4\linewidth]{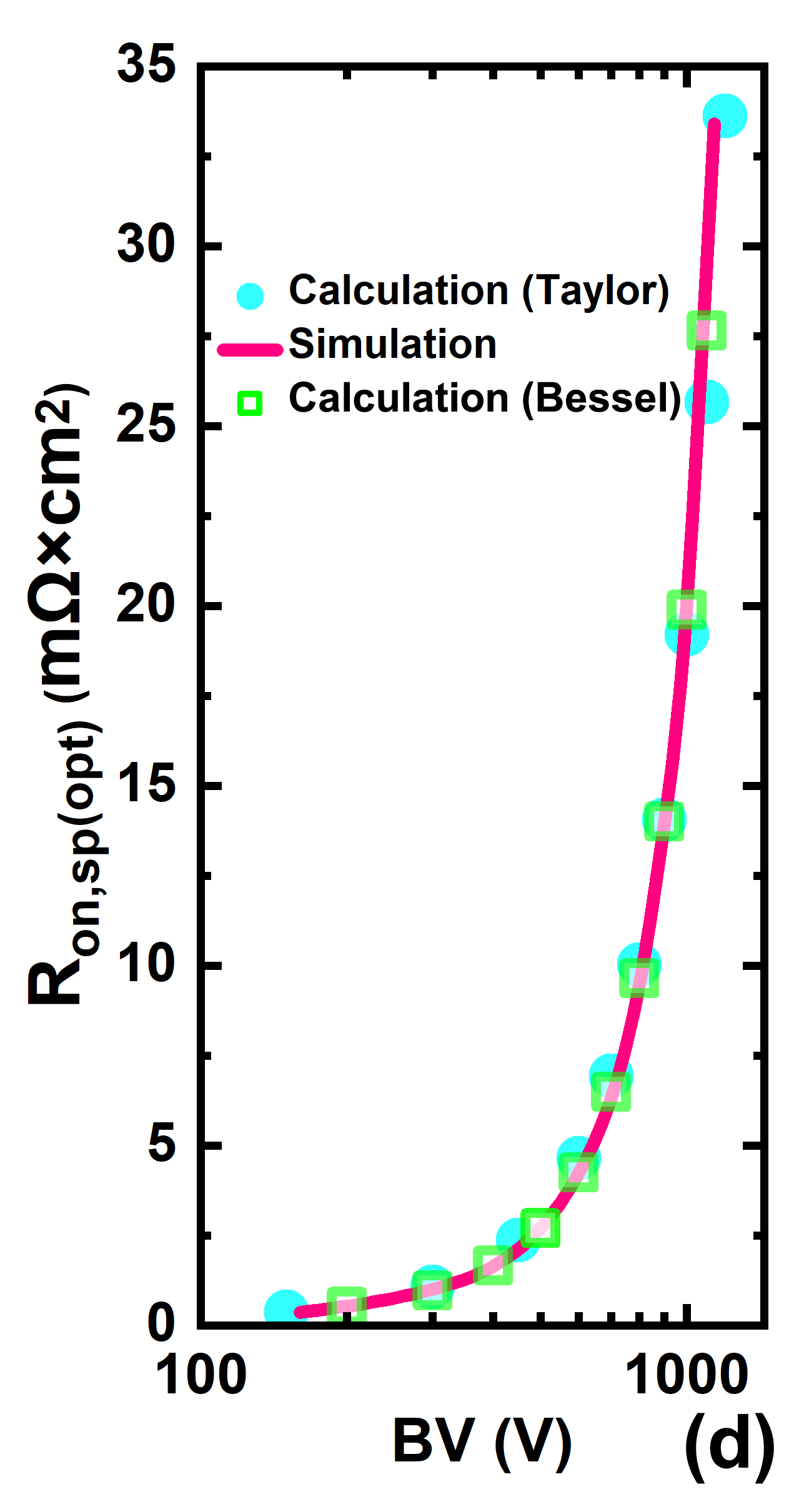} 
    \vspace{-2mm}
    \caption{Aspect ratio dependent optimization results (Taylor method) of (a) $R_{\rm on,sp}$, (b) $N$ and (c) $W$ under the condition of 3DH$k$case2 with a=2 {$\mu$}m and K varies from 20 to 100. (d) $R_{\rm on,sp}$-BV curve used for optimization accuracy confirmation. }
    \label{OPT1}
\end{figure}

\par In fact, we only need to know the expressions for the potential and E-field at r = b, and we do not need to know the complete expressions for $V_{\mathrm{s}}$ and  $V_{\mathrm{H}k}$. This is because, as stated in \cite{2013} and verified through simulations, avalanche breakdown occurs first along the E-field line from $A_{2}$ to $B_{2}$.
\par Fig. \ref{Taylor}(b) shows a comparison of the E-field along $A_{2}$$B_{2}$ for different a/b values. The results obtained using the Taylor method, the Bessel method\cite{2013} and simulation demonstrate good consistency with each other. The error of Taylor method is depicted in Fig. \ref{Taylor}(c) which indicates that with different doping concentration (N), $V_{\rm ap}$ and K (K = $\varepsilon_{\rm S}/ \varepsilon_{\mathrm{H}k}$), the error along the z-direction at r = b mainly less than 2.5\%. The error at z = 0 is relatively large, but this does not affect the accuracy of subsequent calculations because z = 0  corresponds to the point where the E-field reaches its minimum value ( as shown in Fig. \ref{Taylor}(b)). Since the E-field is minimal at this location, it does not influence the accuracy of the subsequent impact ionization integral calculation.

\subsection{Aspect Ratio Dependent Optimization using Taylor Method}
For superjunction devices, there are many optimization methods available. Unlike in \cite{2013}, here we select the critical depletion and critical breakdown as two constraints. Following the methodology outlined in \cite{methodology}, we can perform an \emph{R}$_{\rm \textbf{on,sp}}$ optimization that depends on the aspect ratio (AR$_{\rm S}$, AR$_{\rm S}$ = W/ [2×(Width of N region)]). \emph{R}$_{\rm \textbf{on,sp}}$ for two cases can be determined as 
\begin{equation}
R_{\rm on,sp}=\begin{cases}\frac{W}{q \mu_{\rm n} N}\cdot\frac{b^{2}}{a^{2}}&,\small{\mathrm{for\; 3DH}k \mathrm{case1}},\\
\frac{W}{q \mu_{\rm n} N}\cdot\frac{b^{2}}{b^{2}-a^{2}}&,\small{\mathrm{for\; 3DH}k \mathrm{case2}}.
\end{cases}
\end{equation}

\normalsize{
The breakdown conditions can be determined through the calculation of impact ionization integral. The Chynoweth model \cite{chynoweth1958ionization} is appiled for higher accuracy of confirming breakdown,
}
\begin{equation}
    I_{\rm n}=\int \alpha_{\rm  n} \, \exp \left (  {\int^s (\alpha_{\rm p} -\alpha_{\rm  n}) ds'} \right )  ds.
    \label{in}
\end{equation}

\par Using the methodology from \cite{methodology} and MATLAB calculations, the optimization results for \emph{R}$_{\rm \textbf{on,sp (opt)}}$, \emph{N}$_{\rm \textbf{(opt)}}$ and \emph{W}$_{\rm \textbf{(opt)}}$ are presented in Fig. \ref{OPT1}(a), (b) and (c), respectively, for given values of BV (800 V and 1000 V). From the three figures, it can be observed that under the two constraints mentioned before and given BV, as AR$_{\rm S}$ increases, the doping concentration \emph{N}$_{\rm \textbf{(opt)}}$ gradually increases while the device height \emph{W}$_{\rm \textbf{(opt)}}$ decreases, resulting in a reduction in \emph{R}$_{\rm \textbf{on,sp (opt)}}$. This implies that, in the manufacturing process, one can achieve a lower \emph{R}$_{\rm \textbf{on,sp (opt)}}$ by increasing the AR$_{\rm S}$ (or, say, reducing the width of the N-region). Additionally, using high dielectric constant materials as high-$k$ material and reducing the BV can further reduce \emph{R}$_{\rm \textbf{on,sp (opt)}}$. Specifically, the minimum value, 4.156 $m\Omega \cdot cm^{2}$,  can be achieved when BV = 800 V, K = 100, AR$_{\rm S}$ = 70, \emph{W}$_{\rm \textbf{(opt)}}$ = 50.093 $\mu$m and \emph{N}$_{\rm \textbf{(opt)}}$ = 2.49×$10^{16}$ $cm^{-3}$. Fig. \ref{OPT1}(d) shows the relationship between \emph{R}$_{\rm \textbf{on,sp (opt)}}$ and BV through Taylor method, Bessel method and simulation. The consistency of the three results indicates the high accuracy of above optimization using Taylor method.

\begin{figure}[t]
    \centering
    \includegraphics[width=.45\textwidth]{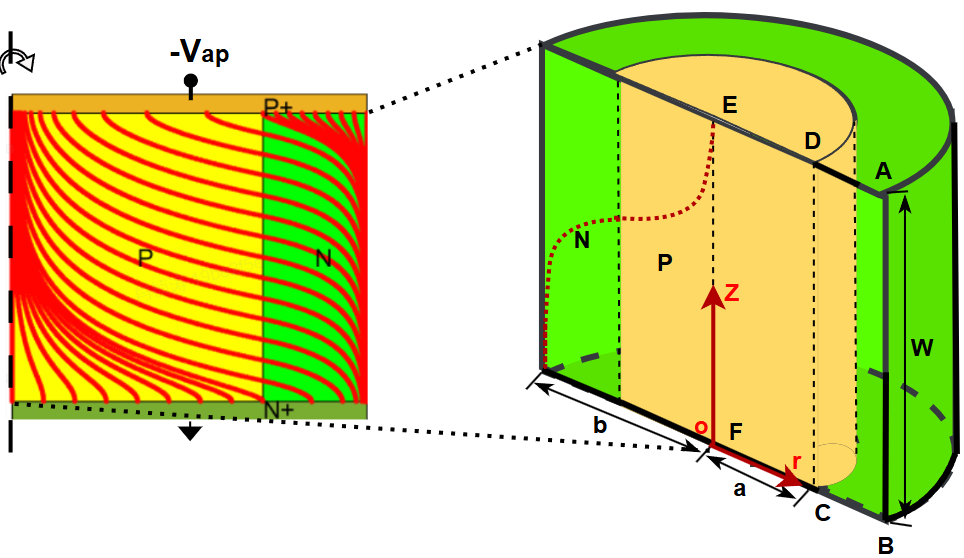}
    \vspace{-3mm}
    \caption{Structure and E-field lines of 3D C-SJ. The E-field lines are calculated and drawn by MATLAB using the methodology in \cite{EFL}. }
    \label{structure2}
\end{figure}

\begin{figure}[t]
    \centering
\includegraphics[width=0.85\linewidth]{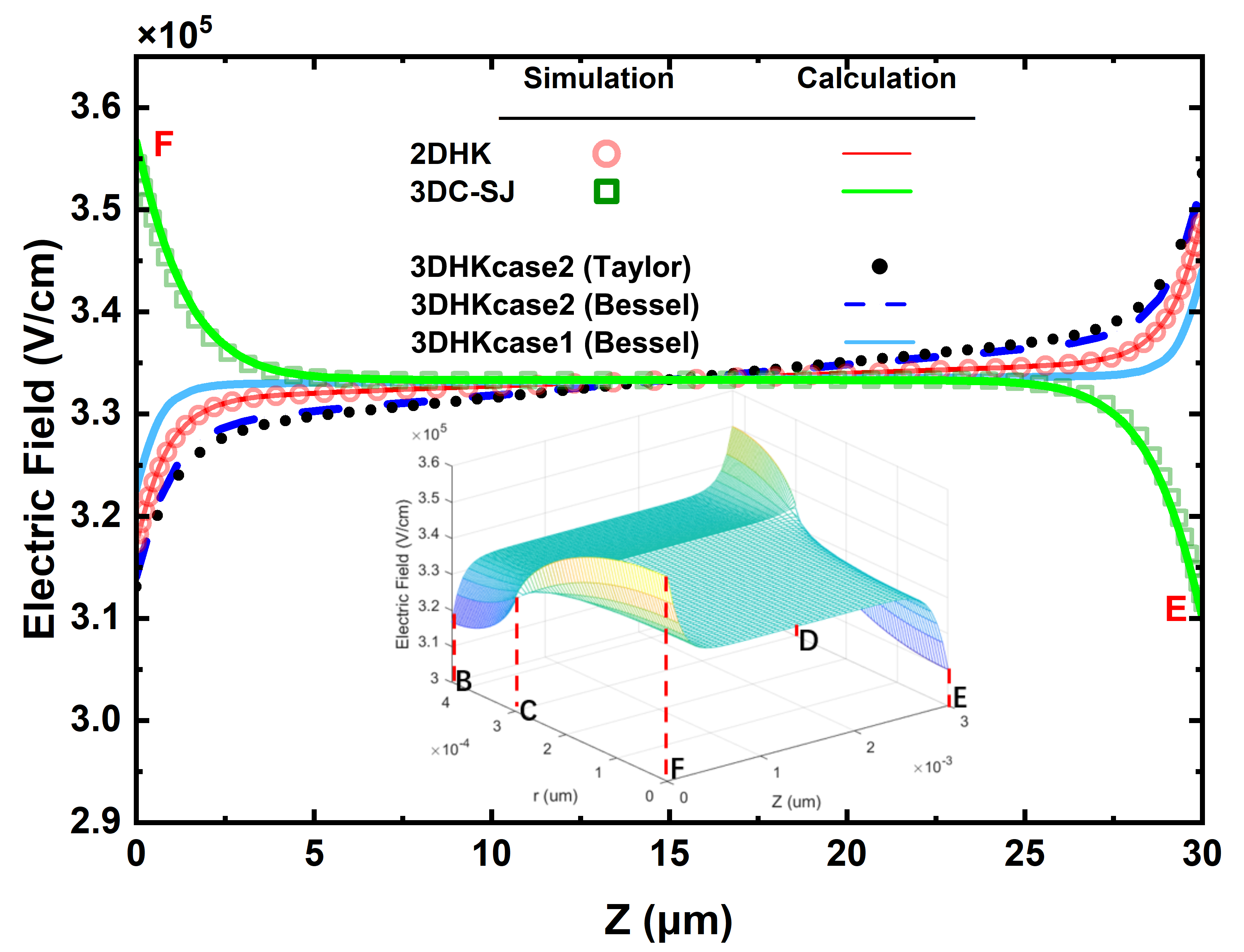}
    \vspace{-3mm}
    \caption{E-field comparison of 4 structures along breakdown path ($E_{1}$$F_{1}$ for 3DH$k$case1 and 2DH$k$, $A_{2}$$B_{2}$ for 3DH$k$case2, $E$$F$ for 3D C-SJ), with 3D depiction of entire E-field of 3D C-SJ.   }
    \label{efield}
\end{figure}

\section{Comprehensive Comparative Analysis of 3DH$k$-SJ}
\label{section3}
 \par Before the comparative analysis, the structure and E-field lines of 3D C-SJ is introduced in Fig. \ref{structure2}. The structure of 2DH$k$ is the same as the profile of 3DH$k$case1. For 3D C-SJ, AR$_{\rm S}$ = W/ [2×(Width of P region)] and ensure charge neutrality for 3D C-SJ in all subsequent discussions.

\subsection{Breakdown E-field Comparative Analysis}

\par With all the parameters set to: $V_{\rm ap}$ = 1000 V, W = 30 $\mu$m, the width of the N region is 1.243 $\mu$m , the width of the P region (H$k$ region) is 3 $\mu$m, K = 50, N = 1×10$^{15}$ cm$^{-3}$, the E-field along the breakdown path for each structure under same condition are shown in Fig. \ref{efield}. It is obvious that, from the figure, 3D C-SJ has the highest peak E-field and 3DH$k$case1 has the lowest, suggesting that the 3D C-SJ is more prone to breakdown under the same conditions, while the 3DH$k$case1 is less likely to break down under the same conditions.

\subsection{Ionization Integral Comparative Analysis}

     \par
\begin{figure}[t]
    \centering
    \includegraphics[width=0.85\linewidth]{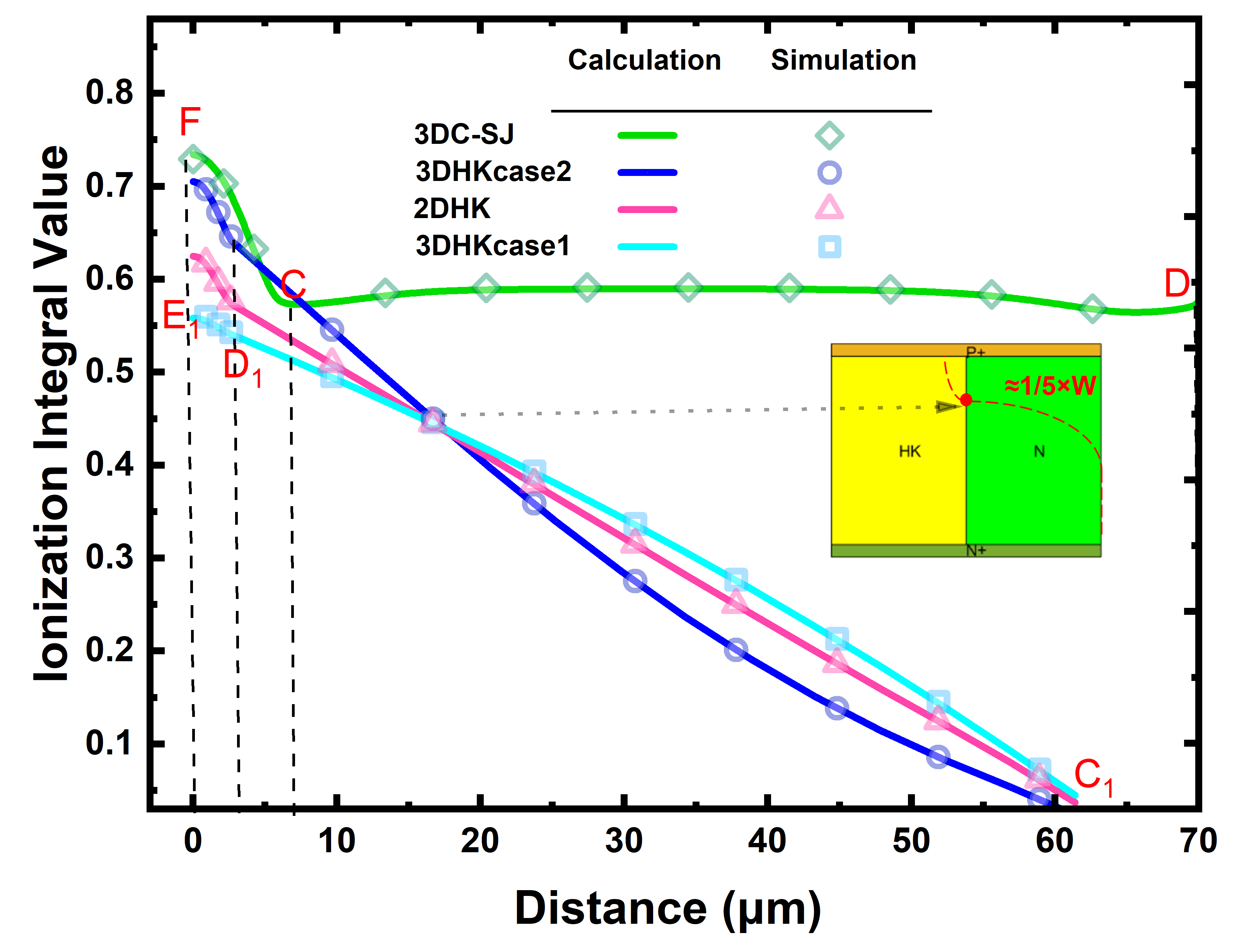}
    \vspace{-3mm}
    \caption{Simulation and calculation results of impact ionization integral values along $E_{1}$$D_{1}$$C_{1}$ for 3DH$K$case1 and 2DH$k$, $A_{2}$$D_{2}$$C_{2}$ for 3DH$k$case2 and $F$$C$$D$ for 3D C-SJ.} 
    \label{ion}
\end{figure}

 To further determine the breakdown characteristics of these structures, the calculation and comparison of the impact ionization integral are necessary. Using equation (10) and methodology in \cite{EFL} with parameters set as follows\cite{hhm202203}: W = 63.3 $\mu$m, the width of the P region (H$k$ region) = 6.33 $\mu$m, the width of the N region = 2.62 $\mu$m, N = 1.97×10$^{15}$ cm$^{-3}$, $V_{\rm ap}$ = 900 V, K = 50. The values of the impact ionization integral along specific paths are shown in Fig. \ref{ion}. From the figure, it can be seen that consistent with the conclusion of Fig. \ref{efield}, the 3D C-SJ has the largest impact ionization integral value at point F, meaning it is more likely to occur avalanche break down under the same conditions, while the 3DH$k$case1 is the least likely to occur.

\subsection{Aspect Ratio Optimization Comparative Analysis}

\begin{figure}[t]
    \centering
    \includegraphics[width=0.85\linewidth]{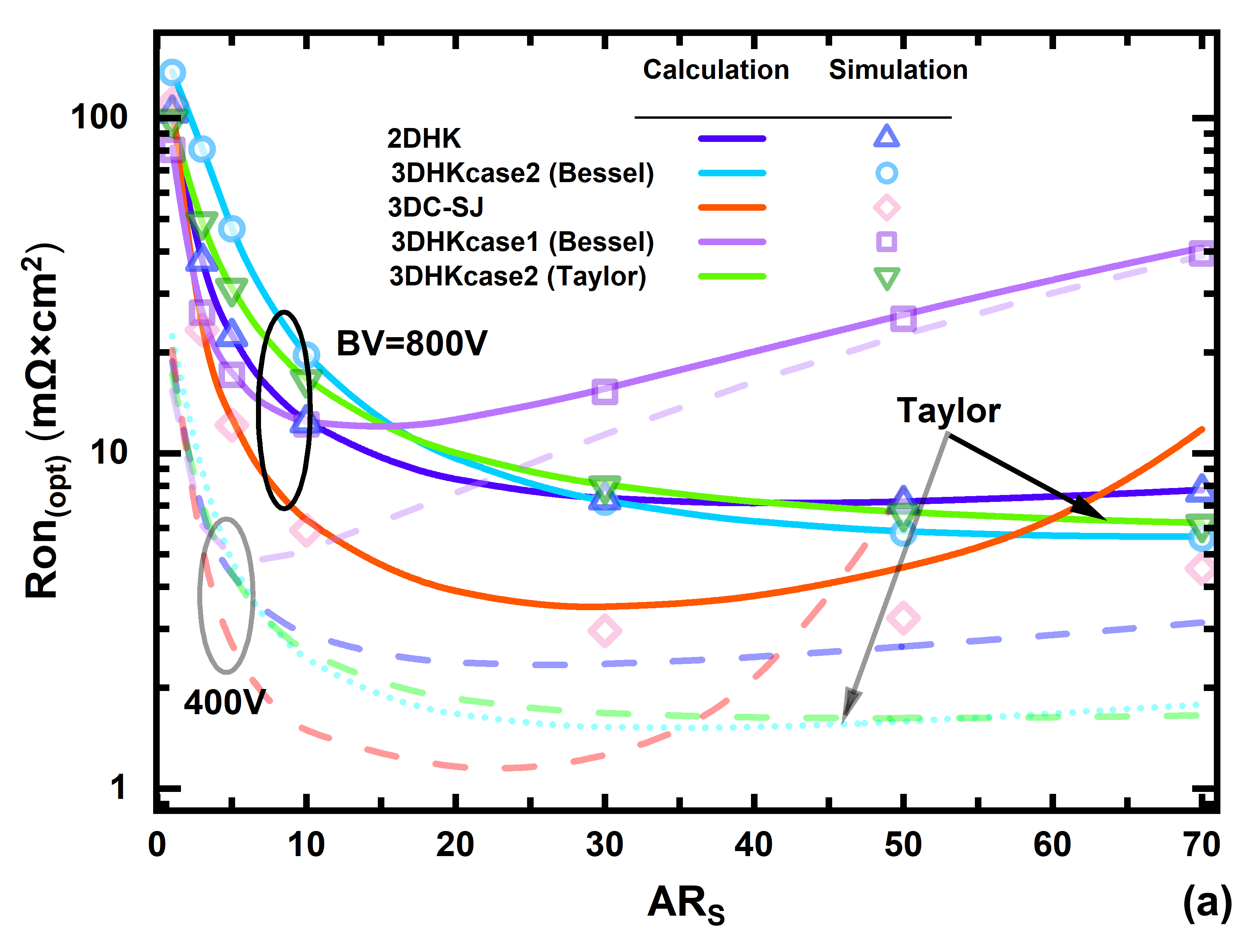}
    \includegraphics[width=0.4\linewidth]{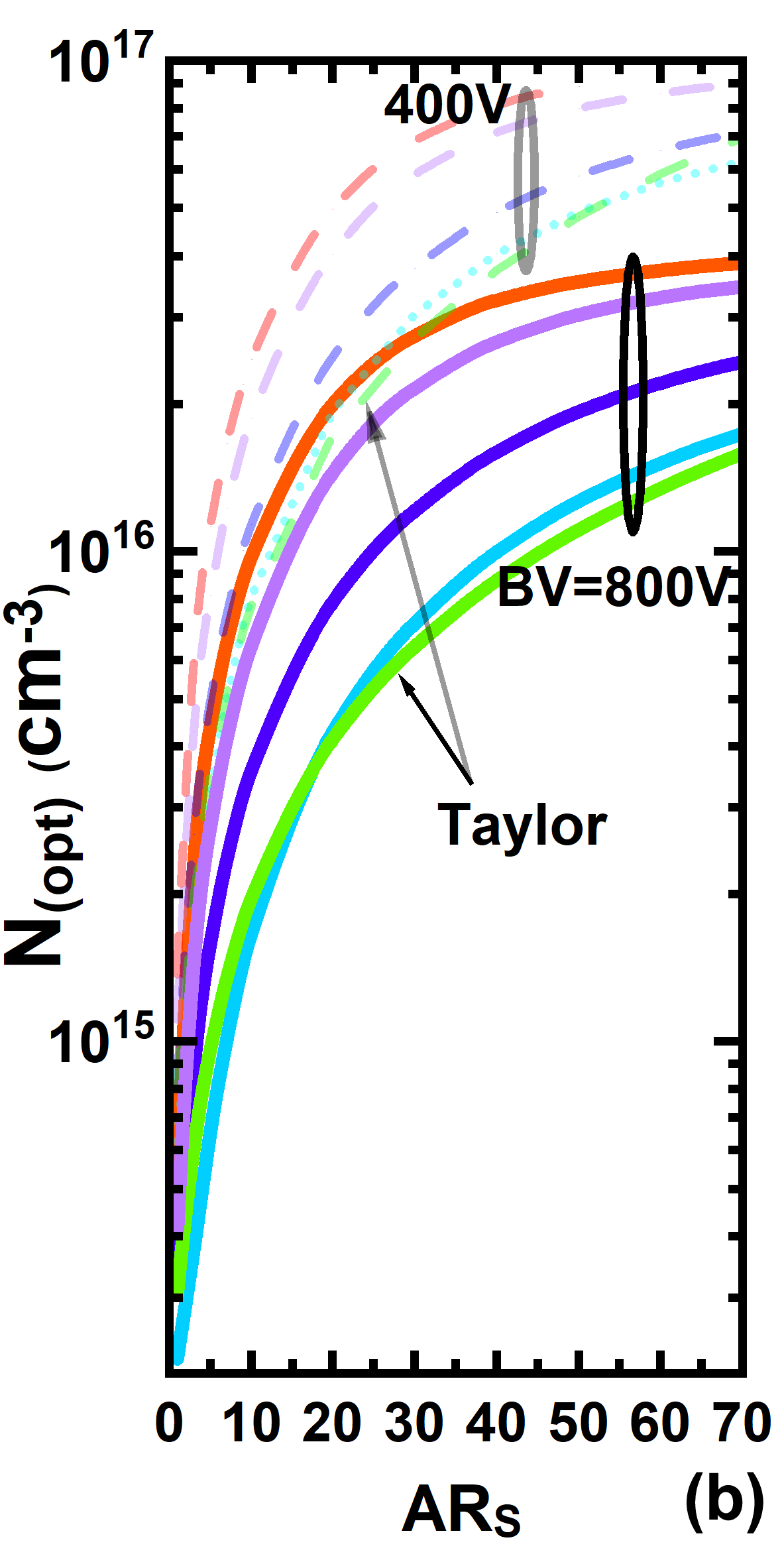}
       \hspace{5mm}
    \includegraphics[width=0.4\linewidth]{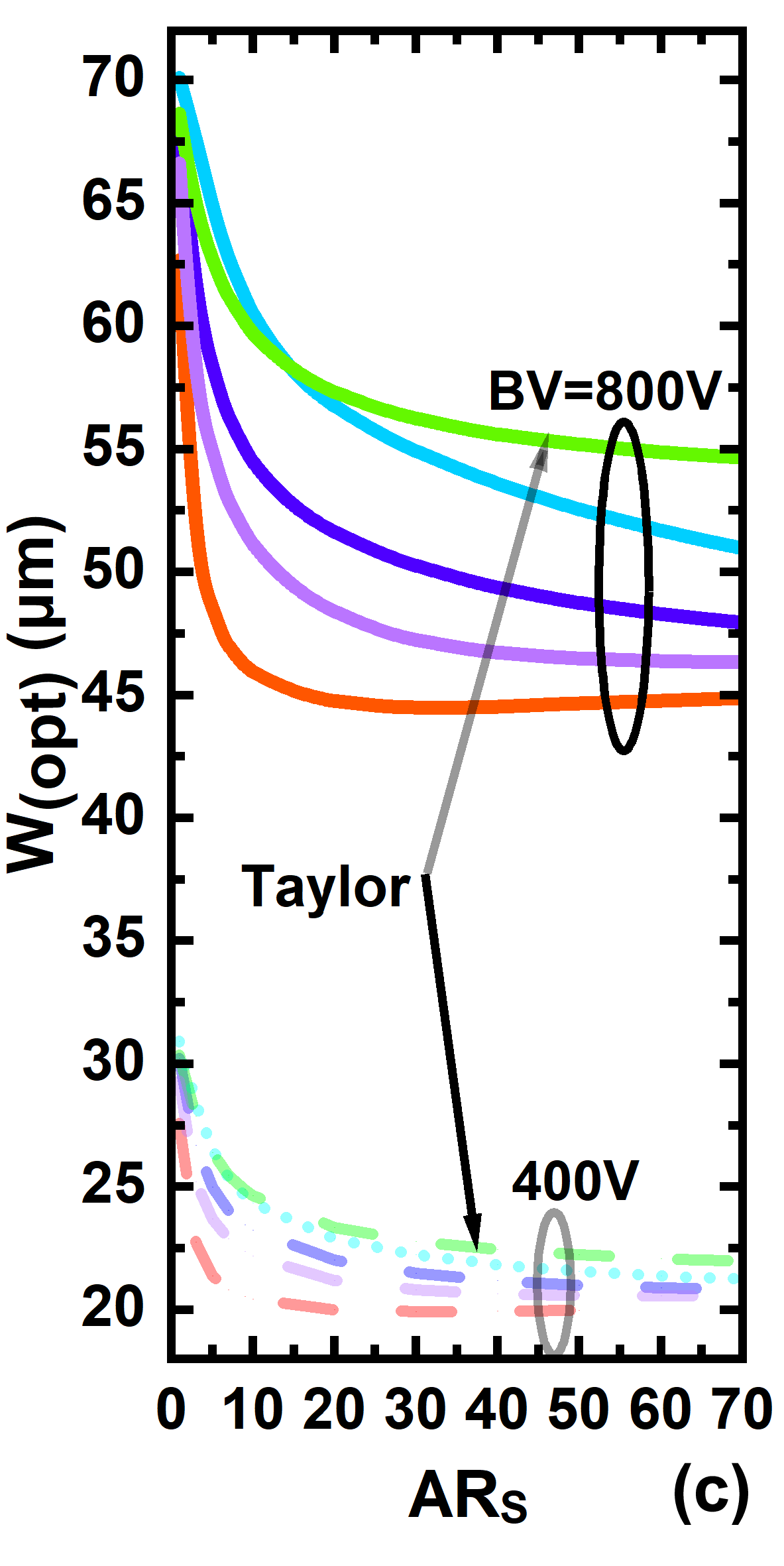}
    \vspace{-2mm}
    \caption{Aspect ratio dependent optimization results for the 4 structures (containing Taylor method) of (a) $R_{\rm on,sp}$, (b) $N$ and (c) $W$ under the condition of H$k$ region width=2 {$\mu$}m, K=50, BV= 400 V and 800 V.}
    \label{OPT2}
\end{figure}

\begin{figure}[!h]
    \centering
     \subfigure[]{\includegraphics[width=0.7\linewidth]{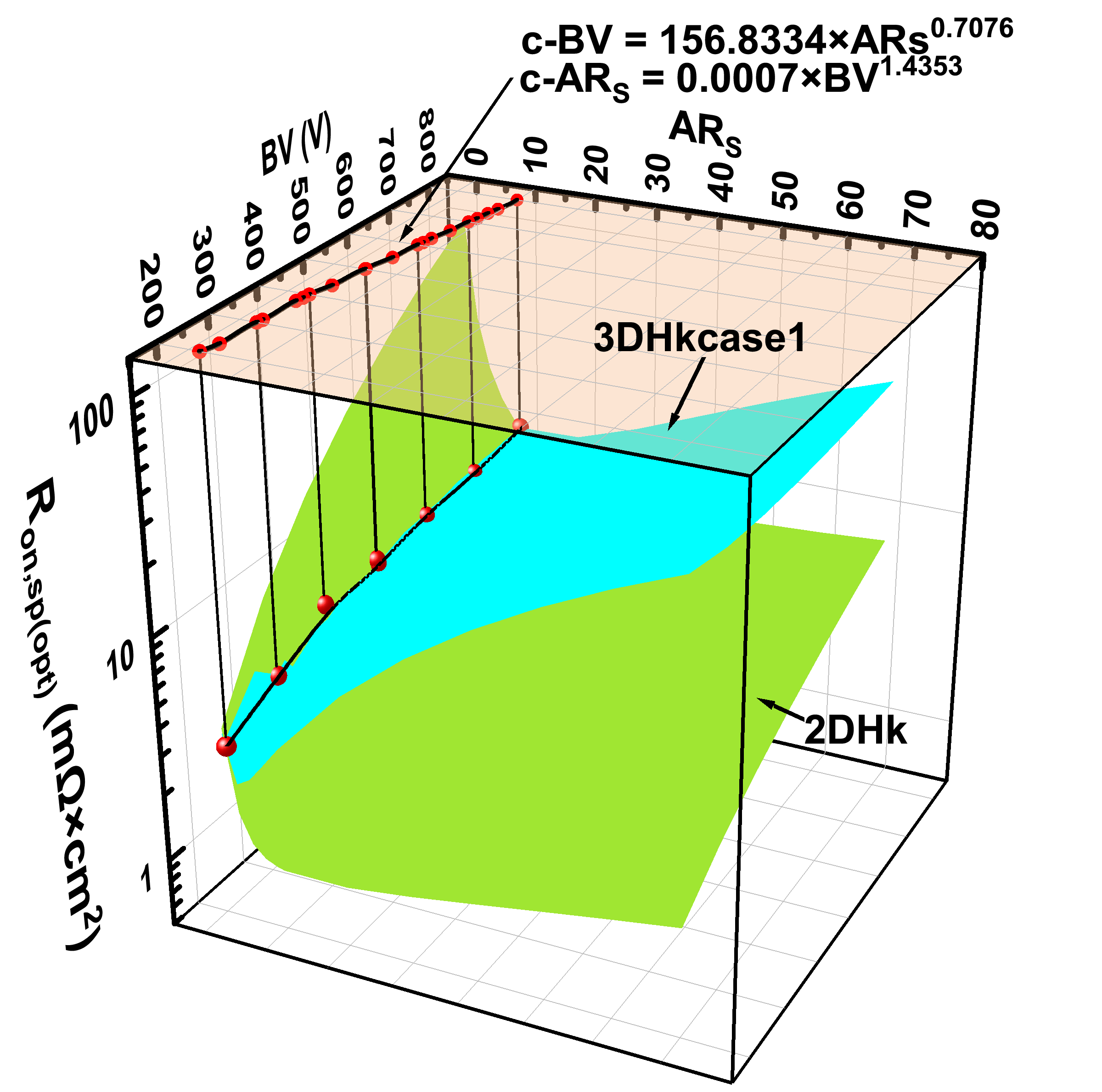}
    }
     \subfigure[]{\includegraphics[width=0.7\linewidth]{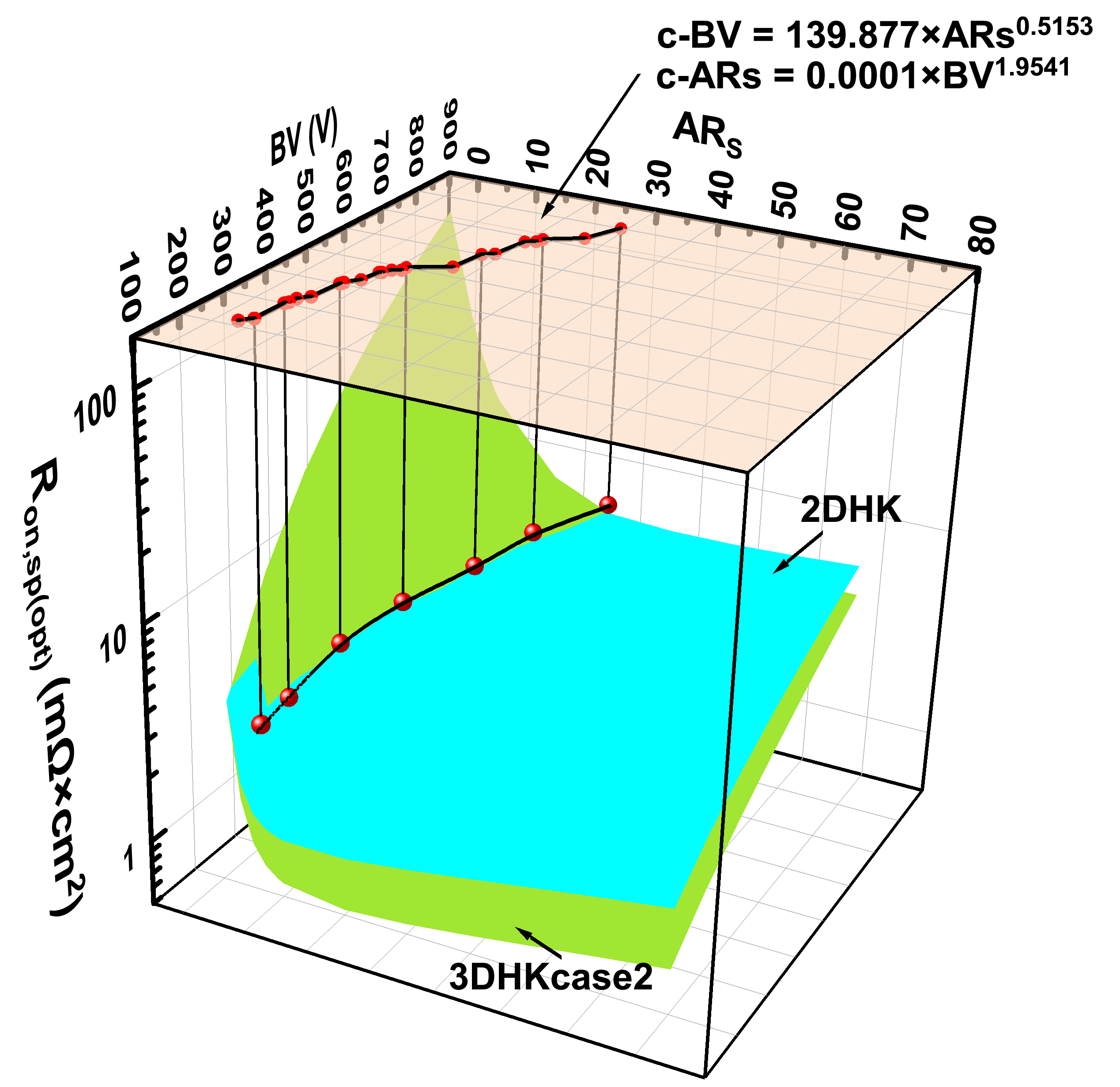}
    }
     \subfigure[]{\includegraphics[width=0.7\linewidth]{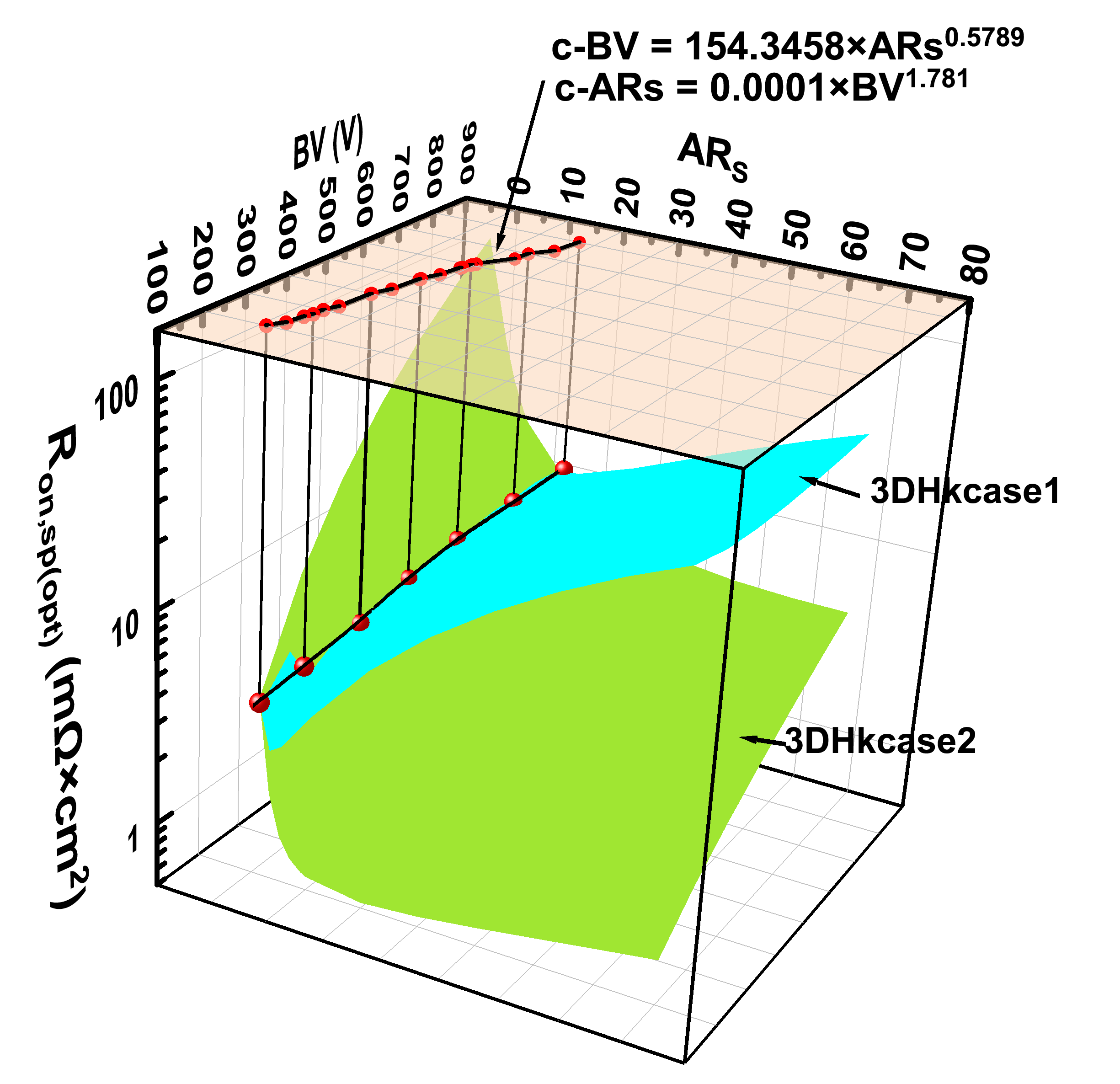}
     }
   \vspace{-2mm}
    \caption{Boundary curves with same \emph{R}$_{\rm \textbf{on,sp (opt)}}$ for (a) 3DH$k$case1 and 2DH$k$, (b) 2DH$k$ and 3DH$k$case2, (c) 3DH$k$case1 and 3DH$k$case2, along with their boundary curves' expressions.}
    \label{3DOPT}
\end{figure}

\begin{table}[b]
\setlength{\tabcolsep}{3pt}
\caption{Optimization Results of 3D C-SJ, 2DH$k$, 3DH$k$case1, and 3DH$k$case2 at BV = 800 V, H$k$ region width = 2 {$\mu$}m and K = 50}
\centering
\begin{tabular}{cccccc}
\toprule[0.5mm]
Symbol&\textbf{3D C-SJ}&\textbf{2DH$k$}&\textbf{3DH$k$case1}&\textbf{3DH$k$case2}&Unit \\
\midrule[0.5mm]
$N_{(opt)}$&2.75×$10^{16}$&1.59×$10^{16}$&6.41×$10^{15}$&1.73×$10^{16}$&$cm^{-3}$ 
\\

$W_{(opt)}$&44.481&49.357&51.137&50.935&{$\mu$}m \\

a&0.741&0.617&2.557&2&{$\mu$}m \\
b&1.048&2.617&4.557&2.364&{$\mu$}m \\
\midrule
$R_{\rm on,sp (opt)}$&3.488&7.104&12.427&5.653&$m\Omega \cdot cm^{2}$ \\
\bottomrule[0.5mm]
\end{tabular}
\end{table}

The optimization trends of the four structures are shown in Fig. \ref{OPT2}(a), (b), and (c). The optimized design parameters acquired from Fig. \ref{OPT2}(a) are shown in TABLE I. For the 3DH$k$case2, the optimization results and trends are similar to those in Fig. \ref{OPT1}. Additionally, the optimization results of the Taylor method and the Bessel method exhibit a high degree of consistency. It can be observed from Fig. \ref{OPT2}(a) and TABLE I that, except for the 3DH$k$case2, the remaining three structures all have a minimum value for \emph{R}$_{\rm \textbf{on,sp (opt)}}$. For the 3DH$k$case1, the minimum value (12.427) is achieved at a small AR$_{\rm S}$, and its minimum value is higher than that of 3D C-SJ, 2DH$k$ and 3DH$k$case2 by 72\%, 43\% and 55\%, implying that the 3DH$k$case1 structure is challenging to reduce \emph{R}$_{\rm \textbf{on,sp (opt)}}$ under the given BV. It is noteworthy that the 3D C-SJ can optimize a smallest \emph{R}$_{\rm \textbf{on,sp (opt)}}$ (3.488) at a larger AR$_{\rm S}$ compared to the other structures, which means that the 3D C-SJ can achieve a more favorable \emph{R}$_{\rm \textbf{on,sp (opt)}}$ under the given BV 
 compared to other structures. For the 2DH$k$, the minimum value is relatively less pronounced, however, its smallest \emph{R}$_{\rm \textbf{on,sp (opt)}}$ (7.104) lies between that of the 3DH$k$case2 (5.653) and the 3DH$k$case1 (12.427). The minimum \emph{R}$_{\rm \textbf{on,sp (opt)}}$ for the 3DH$k$case2 is still achieved at the maximum AR$_{\rm S}$, with its smallest \emph{R}$_{\rm \textbf{on,sp (opt)}}$ (5.653) achieve 120\% and 26\% better than that of the 3DH$k$case1 and 2DH$k$, but 38\% worse than that of 3D C-SJ.
 \par Fig. \ref{3DOPT}(a), (b), and (c) present 3D plots of \emph{R}$_{\rm \textbf{on,sp (opt)}}$ as a function of AR$_{\rm S}$ and BV, from which we can derive guidance for production and manufacturing. Taking Fig. \ref{3DOPT}(a) as an example, for larger AR$_{\rm S}$ and smaller BV, 2DH$k$ can achieve a smaller \emph{R}$_{\rm \textbf{on,sp (opt)}}$, while for smaller AR$_{\rm S}$ and larger BV, 3DH$k$case1 can achieve a smaller \emph{R}$_{\rm \textbf{on,sp (opt)}}$. Quantitatively speaking, if the AR$_{\rm S}$ that can be achieved in device production is determined, the critical BV (c-BV) can be obtained through the boundary curve formula shown in Fig. \ref{3DOPT}(a). If a device which requires a BV greater than the c-BV needs to be produced, 3DH$k$case1 is the better choice because of lower \emph{R}$_{\rm \textbf{on,sp (opt)}}$. If the BV of the device is determined, the critical AR$_{\rm S}$ (c-AR$_{\rm S}$) can be obtained through the boundary curve formula. If the production process supports the creation of a large AR$_{\rm S}$, i.e., AR$_{\rm S}$ $>$ c-AR$_{\rm S}$, 2DH$k$ is the better choice. The same principle applies to Fig. \ref{3DOPT}(b) and (c). By using the formulas of the boundary curves, we can compare the advantages and disadvantages of 2DH$k$, 3DH$k$case1 and 3DH$k$case2 under given conditions, thereby providing valuable guidance for the production of 3D H$k$-SJ devices.

\begin{figure}[h]
    \centering
    \includegraphics[width=0.85\linewidth]{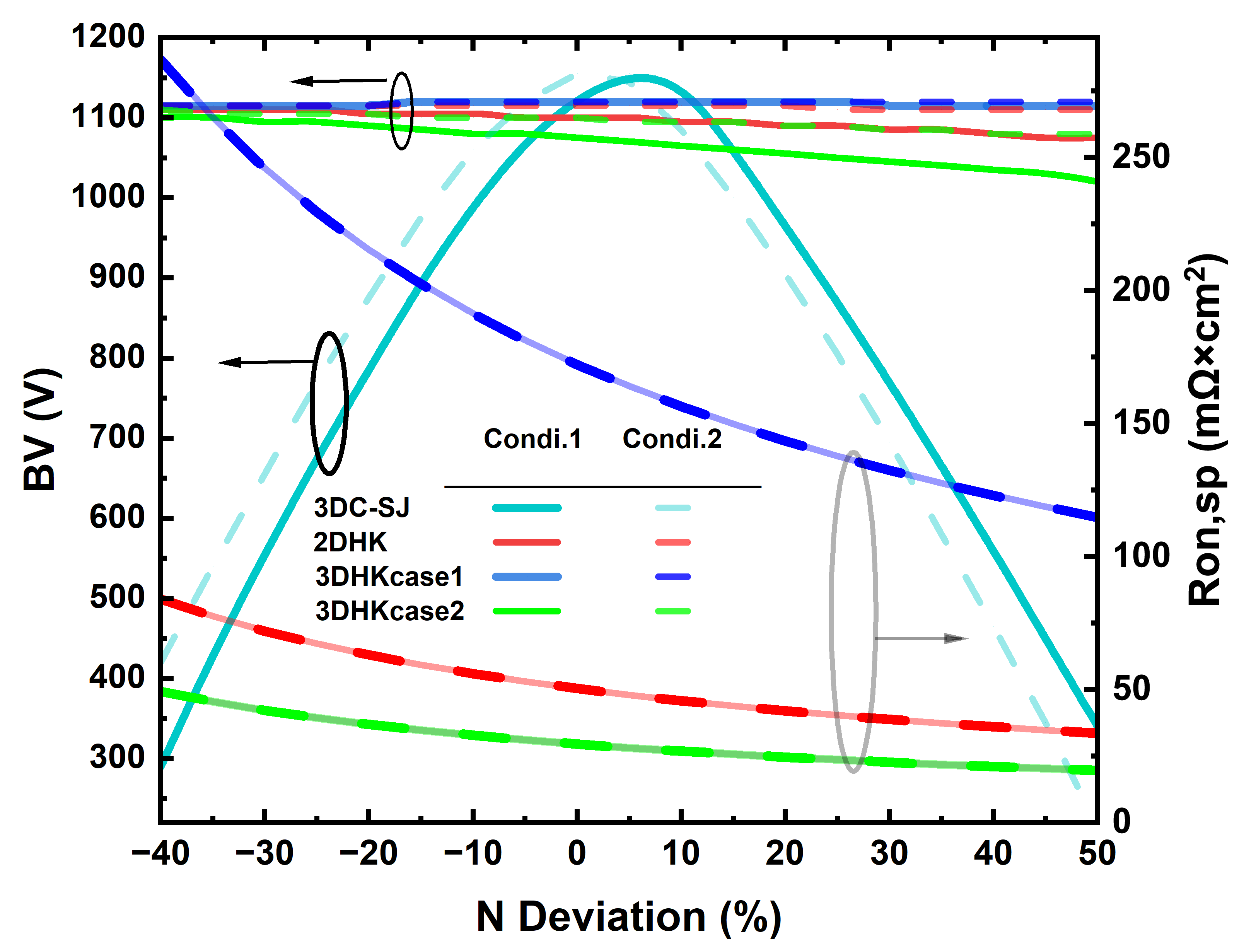}
   \vspace{-2mm}
  
   \caption{Charge imbalance effect of BV and \emph{R}$_{\rm \textbf{on,sp (opt)}}$ as a function of
deviation of N for the four different structures at condition 1 and 2.}
\label{chargeim}
\end{figure}

\begin{figure}[h]
    \centering
     \subfigure[]{\includegraphics[width=0.4\linewidth]{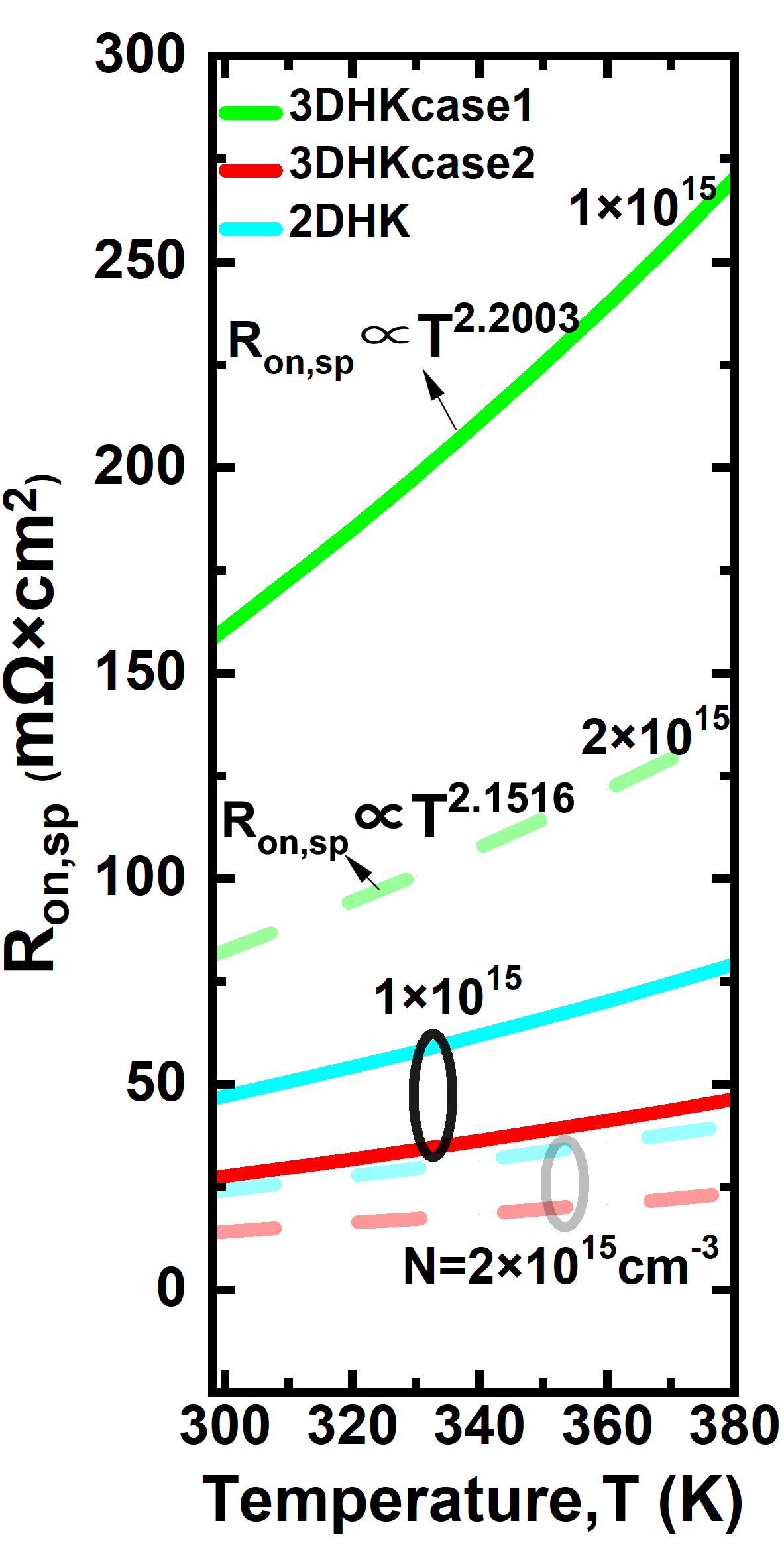}
     \hspace{2mm}
     }
      \subfigure[]{\includegraphics[width=0.4\linewidth]{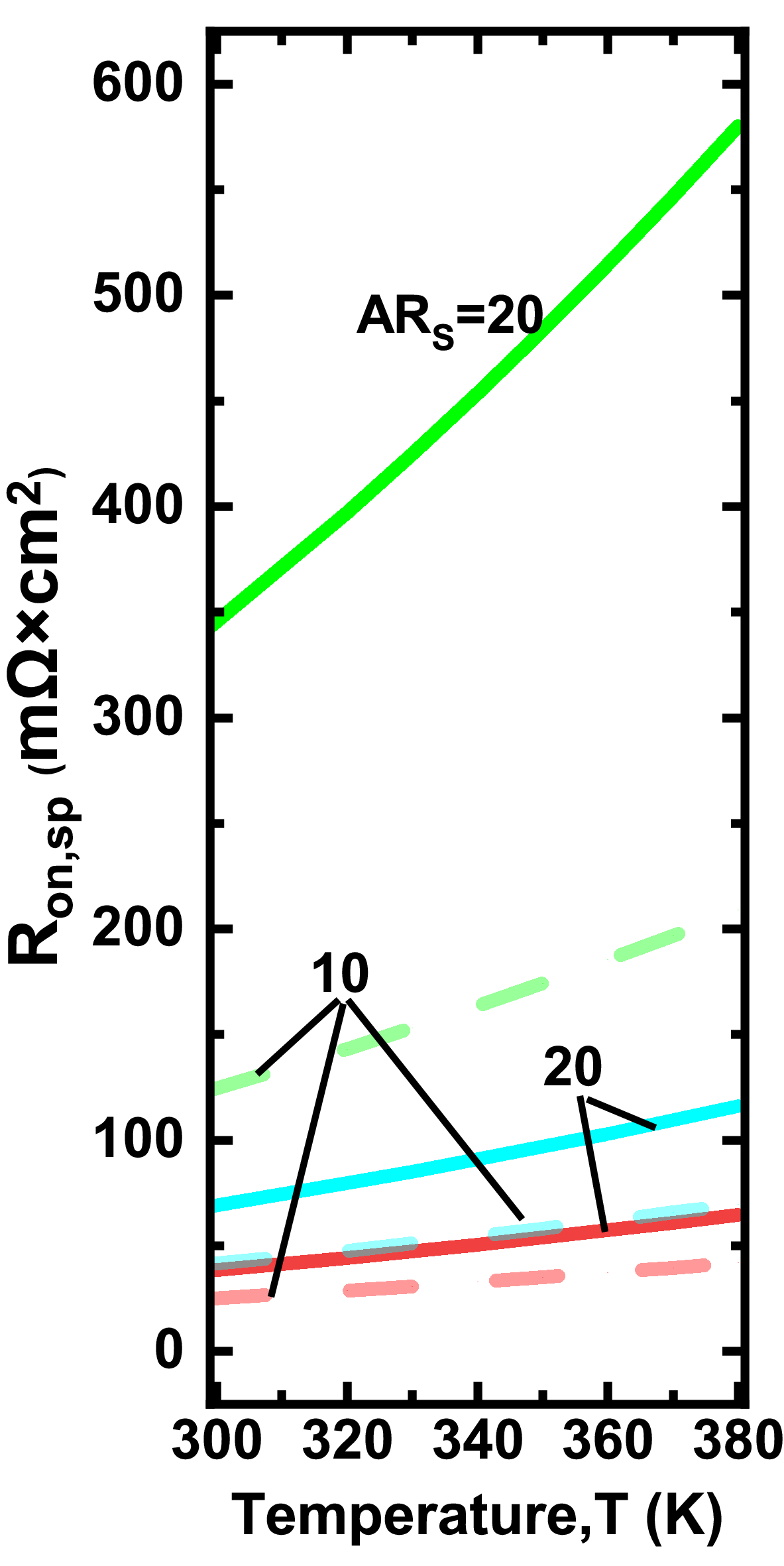}
      }
   \vspace{-3mm}
  
   \caption{\emph{R}$_{\rm \textbf{on,sp}}$ as a function of temperature for three 3D H$k$-SJ structures at (a) different N and (b) different AR$_{\rm S}$. }
   \label{temp}
\end{figure}

\subsection{Charge Imbalance Comparative Analysis}

Fig. \ref{chargeim} illustrates the degradation of BV and \emph{R}$_{\rm \textbf{on,sp (opt)}}$ due to the error in N. The structural parameter settings are divided into two conditions. The first condition (solid lines) is consistent with Fig. \ref{ion}, while the second condition (dashed lines) modifies the width of the P region (H$k$ region) to 3.3 $\mu$m and the width of the N region to 1.367 $\mu$m based on the first condition. It can be observed from the Fig. \ref{chargeim} that regardless of the structure parameters, the sensitivity of BV to errors in all high-$k$ superjunction structures is far less than that of the 3D C-SJ. This implies that the use of various high-$k$ superjunction structures can significantly enhance the robustness of BV against N errors. Regarding \emph{R}$_{\rm \textbf{on,sp (opt)}}$, all H$k$-SJs are noticeably affected by errors, but 3DH$k$case1 is more sensitive to errors compared to 2DH$k$ and 3DH$k$case2.

\subsection{Temperature characteristic Comparative Analysis}
\par Fig. \ref{temp}(a) and (b) depict the temperature dependence of \emph{R}$_{\rm \textbf{on,sp (opt)}}$ for three 3D H$k$-SJ structures under varying $N$ and different AR$_{\rm S}$. From Fig. \ref{temp}(a), it is evident that 3DH$k$case1's \emph{R}$_{\rm \textbf{on,sp (opt)}}$ is the most sensitive to temperature, while the \emph{R}$_{\rm \textbf{on,sp (opt)}}$ of 3DH$k$case2 and 2DH$k$ exhibit similar robustness against temperature variations. This implies that if the device is expected to operate under high-temperature conditions, 3DH$k$case2 and 2DH$k$ are the better choices. Additionally, appropriately increasing the $N$ can reduce the temperature sensitivity of 3DH$k$case1. As shown in Fig. \ref{temp}(a), a higher doping concentration will decrease the exponent of T from 2.2003 to 2.1516.

\begin{figure}[t]
 \centering    \includegraphics[scale=0.39]{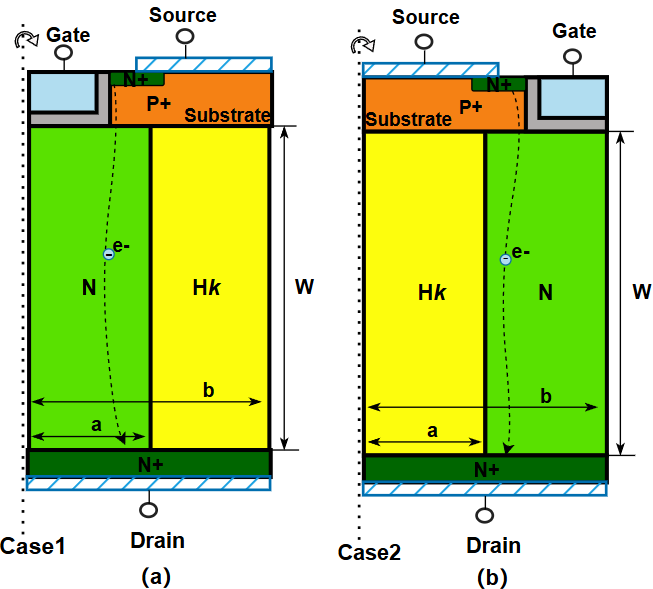}
   \vspace{-2mm}
   \caption{The cross-section structures of (a) 3DH$k$case1 and (b) 3DH$k$case2 MOSFETs. 2DH$k$ MOSFET shares the same structure of 3DH$k$case1.}
   \label{MOS}
\end{figure}

\begin{figure}[!ht]
   \centering
  \includegraphics[width=0.85\linewidth]{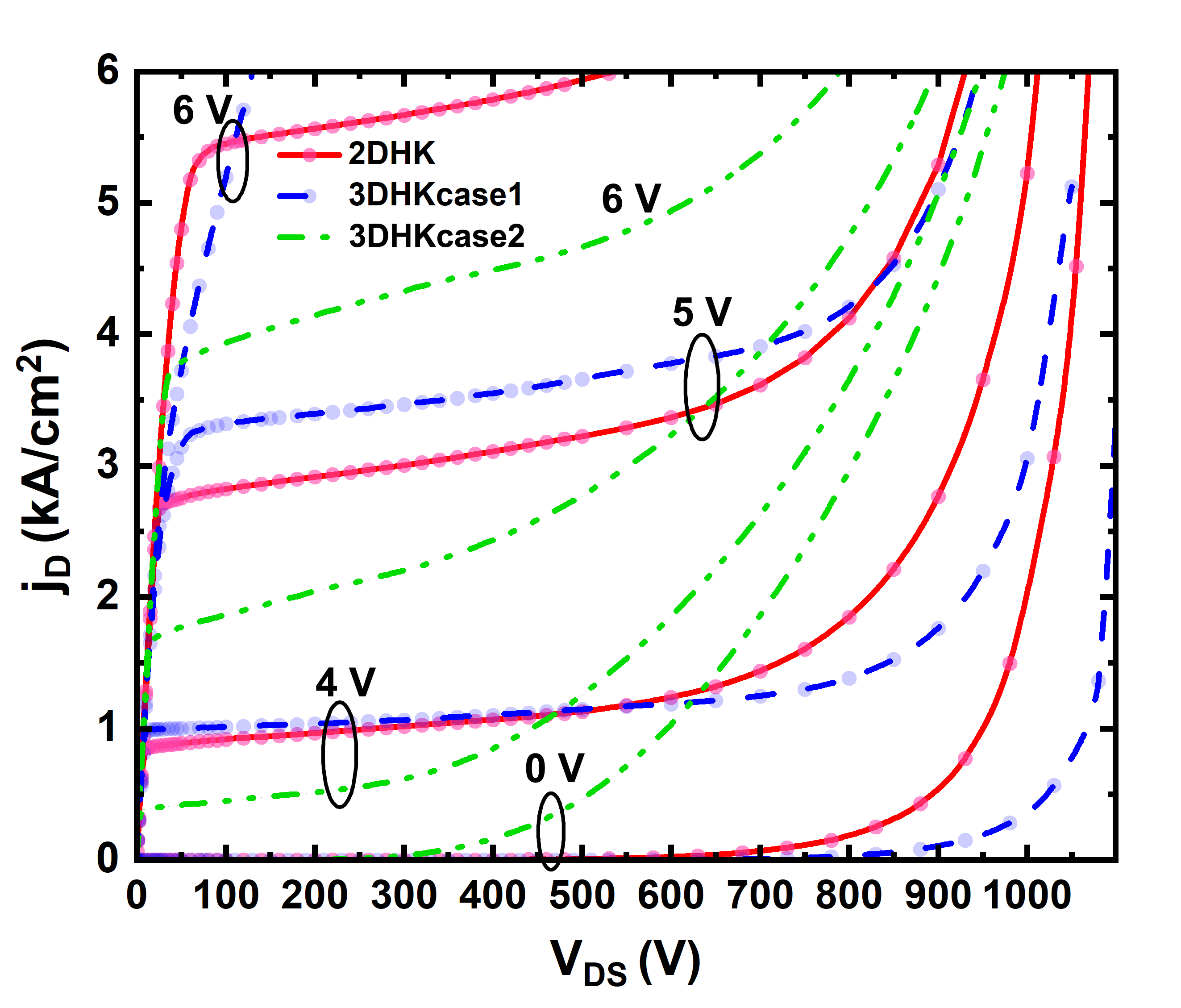}
  \vspace{-3mm}
  
  \caption{Output characteristic of three 3D H$k$-SJ MOSFETs by simulation results at different $V_{\rm GS}$.}
  \label{output}
\end{figure}

\section{Electrical Characteristic Analysis of different H$k$-SJ MOSFETs}

\par The cross-sectional views of the structures for 3DH$k$case1 and 3DH$k$case2 MOSFETs are shown in Fig. \ref{MOS}(a) and (b), respectively, with the meanings of the structural parameters being consistent with Fig. 1. Here, W=75 $\mu$m, N=5×10$^{15}$ cm$^{-3}$, N region width=2 $\mu$m, H$k$ region width=6 $\mu$m, K=20, and the doping concentrations for the Drain region, Substrate, and Source region are 1×10$^{19}$, 2×10$^{17}$, and 3×10$^{19}$ cm$^{-3}$, respectively. The channel length is approximately 4 $\mu$m, and the gate oxide thickness is 35 nm. CONSRH, AUGER, ANALYTIC, FLDMOB, BGN, INCOMPLE and IMPACT.I are used as physical models in MEDICI simulation.

\subsection{Static Output Characteristic}

Fig. \ref{output} demonstrates the static output characteristics and breakdown behavior of three different H$k$-SJ MOSFETs, where $j_{\rm D}$ is the current density of the drain and $V_{\rm DS}$ is the applied voltage across the drain and source. As shown in Fig. \ref{efield} and Fig. \ref{ion}, 3DH$k$case1, with its smaller peak breakdown electric field and impact ionization integral value, is the most difficult one to break down among the three structures at $V_{\rm GS}$ = 0 V, followed by 2DH$k$, while 3DH$k$case2 is the easiest one to break down.

\subsection{Switching Responses Characteristic}

\begin{figure}[!ht]
\centering
    \subfigure[]{\includegraphics[width=0.85\linewidth]{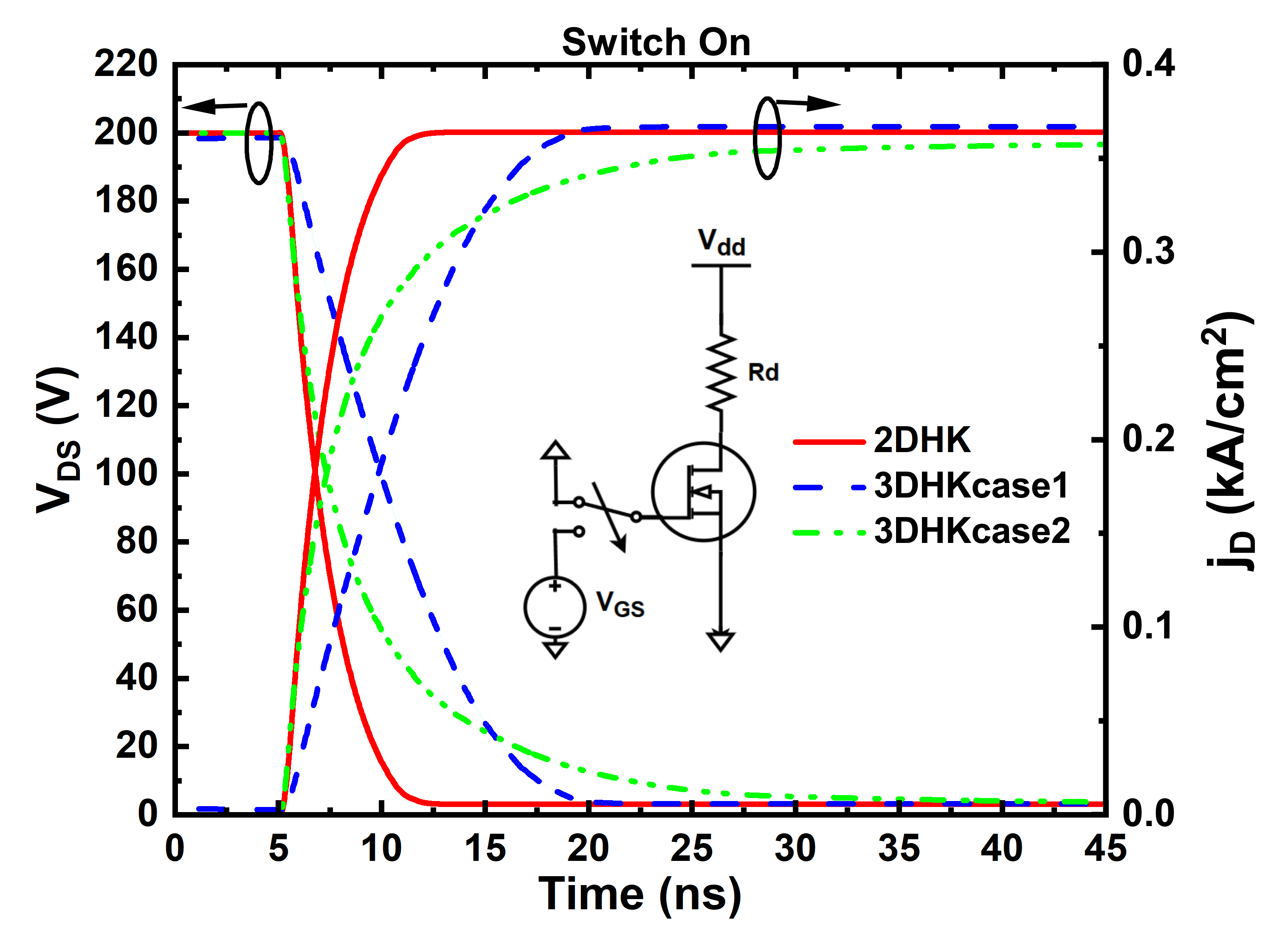}
    }
    \subfigure[]{
\includegraphics[width=0.85\linewidth]{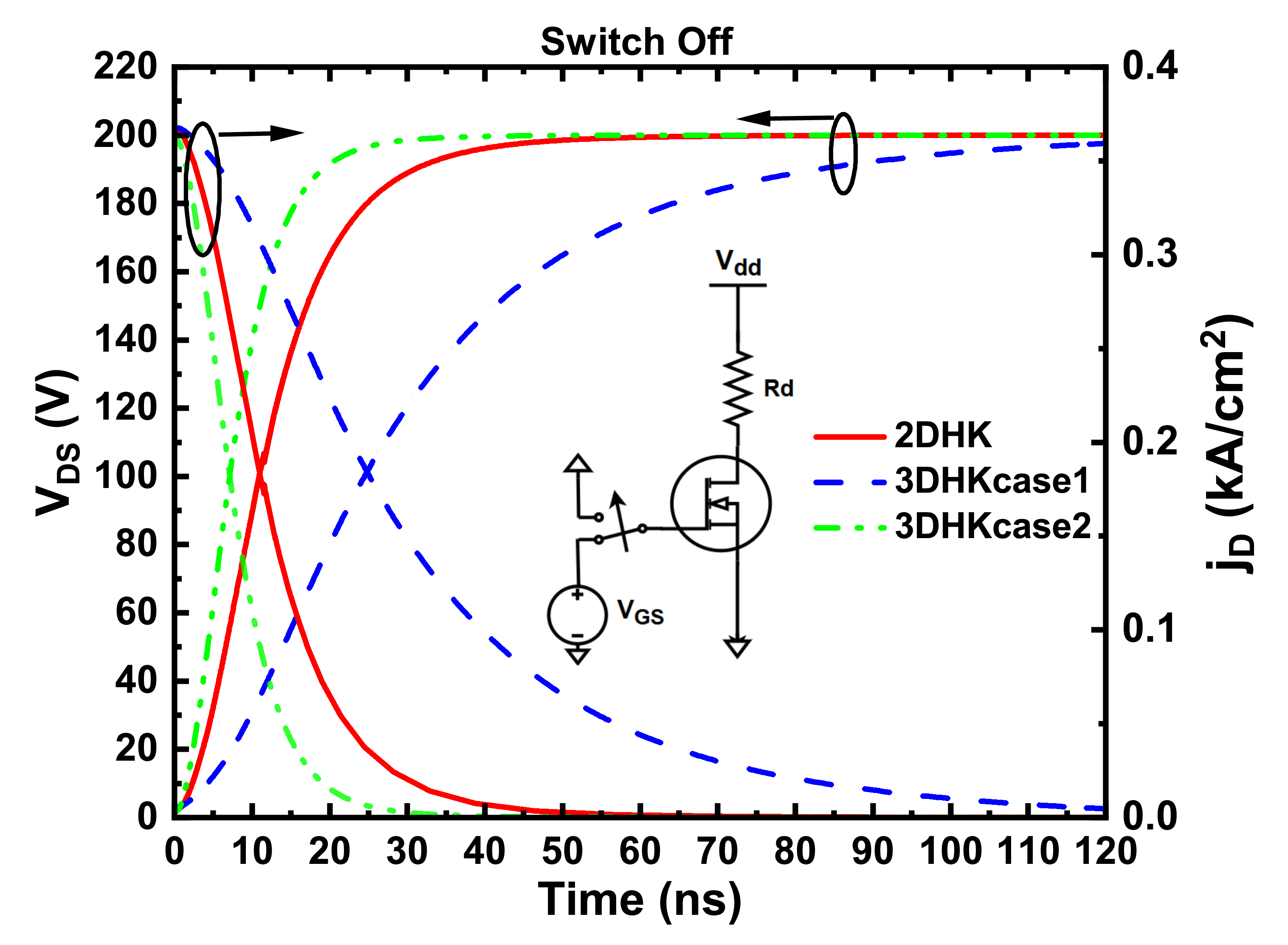}
}
 \vspace{-2mm}
    \caption{ Switching responses of three 3D H$k$-SJ MOSFETs by simulation results. (a) Switch-on. (b) Switch-off}
    \label{sw}
\end{figure}

Fig. \ref{sw}(a) and (b) display the switching responses of the three structures. Here, $V_{\rm GS}$ is ramped up from 0 V to 4 V (and down from 4 V to 0 V) within 0.1 ns, while $V_{\rm dd}$ is maintained at 200 V. From Fig. \ref{sw}(a), it can be observed that 2DH$k$ turns on the fastest, followed by 3DH$k$case1, and then 3DH$k$case2. From Fig. \ref{sw}(b), it is evident that 3DH$k$case2 turns off the quickest, followed by 2DH$k$, and then 3DH$k$case1. This implies that 3DH$k$case1 has relatively poor switching characteristics, while 3DH$k$case2 is suitable for applications in fast chopper circuits.

\section{Conclusion}

\par In summary, this article proposes a Taylor modeling method for 3DH$k$case2 and achieve \emph{R}$_{\rm \textbf{on,sp (opt)}}$ optimization to 4.156 $m\Omega \cdot cm^{2}$ at BV = 800 V and K = 100. Through comparative analysis under the same conditions among 3D C-SJ, 2DH$k$, 3DH$k$case1 and 3DH$k$case2, we find that 3D C-SJ has the potential to optimize for the smallest \emph{R}$_{\rm \textbf{on,sp}}$ to 3.488 $m\Omega \cdot cm^{2}$ at BV = 800 V but is the worst at resisting N deviation. 3DH$k$case1 can have a larger BV compared to 3DH$k$case2 and 2DH$k$, but the \emph{R}$_{\rm \textbf{on,sp (opt)}}$ of 3DH$k$case2 after optimization can be 120\%, 26\% smaller than that of 3DH$k$case1 and 2DH$k$, and case2 also outperforms case1 in terms of switching characteristics and temperature robustness. The performance of 2DH$k$ is usually between case1 and case2. Furthermore, through the formulas of boundary curves, we can provide effective guidance for structure selection in superjunction device manufacturing.


\bibliographystyle{IEEEtran}  
\bibliography{main}           

\end{document}